\documentclass[12pt]{article}
\usepackage{amsmath,amsfonts}
\usepackage{slashed}
\usepackage[dvipdfmx]{graphicx,color}
\usepackage{braket}
\usepackage[nosort]{cite}

\usepackage[shadow,textwidth=2.7cm]{todonotes}

\usepackage{ulem}

\newlength{\extraspace}
\newlength{\extraspaces}
 \catcode`\@=11
\def\numberbysection{\@addtoreset{equation}{section}
\def\theequation{\arabic{section}.\arabic{equation}}}
\newcommand{\nonu}{\nonumber \\[.5mm]}

\newcommand{\tr}{\, \textrm{tr}}

\usepackage{xcolor}
\usepackage[skins,theorems]{tcolorbox}

\tcbset{highlight math style={enhanced,colframe=red,colback=yellow,arc=0pt,boxrule=1pt}}

\usepackage[makeroom]{cancel}

\newcommand{\cL}{{\cal L}}
\newcommand{\cH}{{\cal H}}
\newcommand{\OA}{{\cal{O}(\alpha')}}

\newcommand{\be}{\begin{equation}}
\newcommand{\ee}{\end{equation}}
\newcommand{\bea}{\setlength\arraycolsep{2pt} \begin{eqnarray}}
\newcommand{\eea}{\end{eqnarray}}
\newcommand{\ba}{\begin{array}}
\newcommand{\ea}{\end{array}}
\newcommand{\bn}{\begin{align}}
\newcommand{\en}{\end{align}}

\newcommand{\eq}[1]{\eqref{#1}}
\newcommand{\nn}{\nonumber}
\newcommand{\w}[1]{\\[0.#1cm]}

\newcommand{\bpsi}{{\bar\psi}}

\usepackage{empheq}

\newcommand{\al}{\alpha}
\newcommand{\del}{\delta}
\newcommand{\e}{\epsilon}

\newcommand{\g}{\gamma}

\newcommand{\la}{\lambda}

\newcommand{\m}{\mu}
\newcommand{\n}{\nu}
\newcommand{\om}{\omega}

\newcommand{\pd}{\partial}
\newcommand{\rh}{\rho}
\newcommand{\s}{\sigma}
\newcommand{\ta}{\tau}

\newcommand{\vp}{\varphi}

\newcommand{\PZ}{{\widetilde P}}
\newcommand{\QZ}{{\widetilde Q}}

\newcommand*\circled[1]{\tikz[baseline=(char.base)]{
            \node[shape=circle,draw,inner sep=1.2pt] (char) {#1};}}

\numberwithin{equation}{section}

\begin{document}

\thispagestyle{empty}

{\flushright  MI-HET-763 \\
STUPP-21-250 \\[15mm]}

\begin{center}

{ \Large \bf Dimensional Reduction of Higher Derivative \w2
Heterotic Supergravity}\\[5mm]

\vspace{6mm}
\normalsize
{\large  Hao-Yuan Chang${}^{1}$, Ergin Sezgin${}^{1}$ and Yoshiaki Tanii${}^2$}

\vspace{10mm}

${}^4${\it Mitchell Institute for Fundamental Physics and Astronomy\\ Texas A\&M University
College Station, TX 77843, USA}

\vskip 1 em 

${}^2${\it Division of Material Science, Graduate School of Science and Engineering\\ Saitama University, Saitama 338-8570, Japan}

\vspace{10mm}

\hrule

\vspace{5mm}

\begin{tabular}{p{14cm}}

Higher derivative couplings of hypermultiplets to $6D, N=(1,0)$ supergravity are obtained from dimensional reduction of 10D heterotic supergravity that includes order $\alpha'$ higher derivative corrections. Reduction on $T^4$ is followed by a consistent truncation. In the resulting action the hyperscalar fields parametrize the coset $SO(4,4)/(SO(4)\times SO(4))$. While the $SO(4,4)$ symmetry is ensured by Sen's construction based on string field theory, its emergence at the field theory level is a nontrivial phenomenon. A number of field redefinitions in the hypermultiplet sector are required to remove several terms that break the $SO(4)\times SO(4)$ down to its $SO(4)$ diagonal subgroup in the action and the supersymmetry transformation rules. Working with the Lorentz Chern-Simons term modified 3-form field strength, where the spin connection has the 3-form field strength as torsion, is shown to simplify considerably the dimensional reduction.

\end{tabular}

\vspace{6mm}
\hrule
\end{center}

\newpage

\setcounter{tocdepth}{2}

\tableofcontents

\medskip

\newpage

\section{Introduction}
%%%%%%%%%%%%%%%%%%%%%%%%%%%%

Studies of higher derivative supergravities in lower than ten dimensions with no known string theory origin have uses in exploring whether they may provide effective field theories that may possibly have a consistent UV completion \cite{Vafa:2005ui}. Matter coupled $N=(1,0), 6D$ supergravities\cite{Nishino:1986dc,Nishino:1997ff,Riccioni:2001bg} provide a rich landscape to investigate this question (see, for example, \cite{Kim:2019vuc,Tarazi:2021duw}). In particular, $R$-symmetry gauged and remarkably anomaly free such supergravities exist \cite{RandjbarDaemi:1985wc,Avramis:2005qt,Avramis:2005hc,Pang:2020rir} that are not embedded in string theory, and as such their higher derivative corrections are of great interest. 
At the level of two derivatives, such supergravities have been known for sometime \cite{Nishino:1986dc,Nishino:1997ff,Riccioni:2001bg} and one of their salient features is the occurrence of quaternionic Kahler sigma models that describe the hypermultiplet scalars. Their higher derivative extensions, on the other hand, have not been investigated so far, with the exception of \cite{Bergshoeff:2012ax}, where, however, the hypermultiplet couplings were not considered. One of the motivations for the current work is to initiate this program. We shall not consider $R$-symmetry gauging and Yang-Mills coupling in this paper but we shall study the higher derivative couplings of the hypermultiplets as a first step. We will work on-shell\footnote{Four-derivative $N=(1,0)$ invariants have been constructed off-shell \cite{Bergshoeff:1986vy,Bergshoeff:1986wc,Nishino:1986da,Bergshoeff:1987rb,Novak:2017wqc,Butter:2018wss} but they do not include hypermultiplets. Moreover, the  elimination auxiliary fields gives rise to infinitely many terms whose relation to Noether procedure construction of higher derivative couplings, or string theory low energy effective action, is not entirely clear. The role of highly nonlinear field redefinitions is another complicating factor. For an earlier preliminary work on on-shell $6D$ higher derivative supergravity see \cite{Han:1985pp}.}.

One approach to study of higher derivative extension of matter coupled supergravities is to employ Noether procedure. However, already at the four-derivative level, even with the assumption that the quaternionic Kahler structure is preserved in the case of $N=(1,0), 6D$ supergravity, one finds that an appropriate ansatz contains a large number of terms, and their variations under supersymmetry gives even larger set of structures that need to vanish. Furthermore, it is not guaranteed that the quaternionic Kahler structure can be maintained. One exception is the case of Grassmannian coset $Gr(n,4)=SO(n,4)/(SO(n)\times SO(4))$. It has been proven by Sen \cite{Sen:1991zi} that the dimensional reduction of heterotic supergravity with gauge fields truncated to the Cartan subalgebra must exhibit at string tree level, and therefore to all orders in $\alpha'$, a continuous $O(d, d + 16;R)$ global symmetry, related to the $O(d, d + 16;Z)$ T-duality of heterotic strings on a $d$-torus. See also \cite{Hohm:2014sxa} where the symmetries of S-matrix elements of massless states were used to explain this symmetry. At the two-derivative level, and in the bosonic sector, sometime ago Maharana and Schwarz \cite{Maharana:1992my} showed that reduction on $T^d$ does give an $O(d, d + 16;R)$ invariant result. In a relatively recent work, it was shown that the effective action for the bosonic string, as well as the bosonic sector of the heterotic string at the four-derivative level, in the absence of Yang-Mills fields, do yield $O(d, d;R)$ invariant action upon reduction on $T^d$ \cite{Eloy:2020dko}. Soon after, the Yang-Mills were taken into account to obtain $O(d, d+16;R)$ invariant result \cite{Ortin:2020xdm}, where, however, the fermionic sector was not considered. For an earlier work where only the scalar fields are kept, see \cite{Godazgar:2013bja}. As for the reduction of Type II string effective actions on $K3$ in which only the NS-NS sector fields $(g_{\m\n}, B_{\m\n}, \varphi)$ are kept at the four-derivative level in $6D$, see for example \cite{Liu:2019ses}.
Another approach to obtaining the higher derivative extended $O(d,d)$ invariant supergravities, or their bosonic sector thereof, is to employ the $\alpha'$ extended double field theories \cite{Hohm:2013jaa,Hohm:2014xsa,Marques:2015vua,Baron:2017dvb,Lescano:2021guc}. The reduction of double field theory in the the bosonic sector has been carried out in \cite{Baron:2017dvb}, and we shall comment further on this  in Section 6.
 
Inclusion of the fermionic sector in the reduction requires that the dimension of the torus is specified. In this paper we will work out the dimensional reduction of full heterotic supergravity to six-dimensions, including its fermionic sector, with its order four-derivative order $\alpha'$ corrections a la Bergshoeff and de Roo \cite{Bergshoeff:1989de}, but leaving out the Yang-Mills multiplets, and consistently truncating to $(1,0)$ supersymmetry. 
While the $T^4$ reduction gives $(1,1)$ supergravity multiplet coupled to four $(1,1)$ vector multiplets, the truncation sets to zero the vector fields, and appropriate fermions, resulting in (reducible) $(1,0)$ supergravity, consisting of pure $(1,0)$ supergravity plus a single tensor multiplet, coupled to four $(1,0)$ hypermultiplets. 
As expected, we do find an $O(4,4)$ invariant result in $6D$. More specifically, the hyperscalars parametrize the coset space $SO(4,4)/(SO(4)_+ \times SO(4)_-)$. In arriving at this result, we shall see that there are several terms that naively arise which are invariant only under the $SO(4)$ diagonal subgroup of $SO(4)_+\times SO(4)_-$, and that the required cancellation of all of these terms is nontrivial, requiring elaborate field redefinitions of hyperscalars and hyperfermions. In the computation of the $\OA$ terms in the action and supertransformations, we work from the outset with the Lorentz Chern-Simons modified field strength in which the spin connection has bosonic torsion furnished by the 3-form field strength itself. This approach is shown to simplify the calculations considerably. 
In particular the extension of the Lorentz Chern-Simons term modified 3-form field strength to include a Chern-Simons form built out of the composite connection arises readily.

Given the motivation for the higher derivative extension of supergravities with no known string origin, the reasons for studying the reduction of heterotic supergravity are two-folds. Firstly, once we get a handle on the structure of the higher derivative couplings for the Grassmannian coset $Gr(4,4)$, we expect that it can be extended readily to $Gr(n,4)$ and more to the point, we can deform the theory by $R$-symmetry gauging in an anomaly free fashion. Such extensions typically do not follow from string theory. Second, the lessons learned from the $Gr(n,4)$ case may be utilized in the direct $6D$ construction of higher derivative couplings of the other quaternionic Kahler spaces \cite{Bagger:1983tt,Alekseevsky'75,deWit:1991nm,LeBrun:1991}. Such couplings, unlike the case of $Gr(n,4)$, are not guaranteed, and they will be treated elsewhere. 

The paper is organized as follows. In Section 2 we present the heterotic supergravity action with its four-derivative extension a la Bergshoeff and de Roo. In Section 3, we provide the set up and useful results in working out the dimensional reduction. In Section 4 we obtain the reduction at the two-derivative level, and in Section 5 we carry out the reduction at $\OA$. In Section 6, the field redefinitions as well as the resulting $SO(4,4)$ invariant action, and the reduction of the supertransformations at $\OA$ are given. In Section 6, the bosonic sector of our results are examined more closely, and are shown to agree completely with those of \cite{Eloy:2020dko} and \cite{Baron:2017dvb}, while apparently differing from those of \cite{Ortin:2020xdm}. Our results are summarized and future directions are pointed out in Section 7. Our notations and conventions are given in Appendix A, the field redefinitions are described in  Appendix B, and the 6D action and supertransformations are summarized in their simplest form in Appendix C. 

%%%%%%%%%%%%%%%%%%%%%%%%%%%%%%%%%%%%%%%%%%%%%%%%%%%%%%%%%%%%%%%%%%%%%%%
\section{Higher derivative heterotic supergravity}
%%%%%%%%%%%%%%%%%%%%%%%%%%%%%%%%%%%%%%%%%%%%%%%%%%%%%%%%%%%%%%%%%%%%%%%

The heterotic supergravity multiplet consists of the fields
\be
(\,e_\mu{}^r,\ \psi_\mu\,\ B_{\mu\nu},\ \chi,\ \varphi\,)\ ,
\ee
where the spinors are Majorana-Weyl with chiralities $\gamma_{11}\psi_\mu=\psi_\mu$ and $\gamma_{11} \chi=-\chi$, and $\mu, r = 0,1,...,9$. The Bergshoeff-de Roo extended heterotic supergravity Lagrangian, in the absence of Yang-Mills multiplets, and in string frame and up to quartic fermion terms, takes the form \cite{Bergshoeff:1989de}
\allowdisplaybreaks{
\begin{align}
\cL =& \cL_0 + \mathcal{L}_{0, \, \mathcal{O}(\alpha')} +\cL_{\alpha'}(R^2)\ ,
\label{Lag}
\w2
\mathcal{L}_0 &=ee^{2\vp} \biggl[ 
\tfrac{1}{4} R(\omega) + g^{\m\n} \partial_\m \vp \partial_\n \vp
- \tfrac{1}{12} H_{\m\n\rh} H^{\m\n\rh} \nonu
& \quad  - \tfrac{1}{2} \bpsi_\m \gamma^{\m\n\rh} D_\n(\omega) \psi_\rho 
+ 2 {\bar\chi}\gamma^{\m\n}  D_\m(\omega) \psi_\n
+ 2 {\bar\chi} \gamma^\m D_\m(\omega)\chi \nonu
& \quad  - \tfrac{1}{24} H_{\m\n\rh} \Big( 
\bpsi^\s \gamma_{[\s} \gamma^{\m\n\rh} \gamma_{\tau]}\psi^\tau
+4 \bpsi_\s \gamma^{\s\m\n\rh}\chi
- 4 {\bar\chi} \gamma^{\m\n\rh} \chi \Big) \nonu
&\quad - \partial_\m \vp\Bigl(  \bpsi^\m \gamma^\n \psi_\n + 2\bpsi_\n \gamma^\m 
\gamma^\n \chi\Bigr)  \biggr]\ ,
\label{het}
\w2
\mathcal{L}_{0, \, \mathcal{O}(\alpha)}
&=  \alpha'\,e e^{2\vp} \biggl[   H^{\m\n\rh} \omega^L_{\m\n\rh}(\Omega_-)-H^{\m\n\rh} R_{\m\n}{}^{rs} (\Omega_-)\bpsi_r \gamma_\rh \psi_s
+\bpsi_r\gamma_\n\psi_s\,\Omega_{-\rh}{}^{rs}   \e^{-2\vp} D_\m(\Gamma) \big(e^{2\vp}H^{\m\n\rh} \big)
\nn
\w2
& \quad + \tfrac14\omega^L_{\m\n\rh}(\Omega_-) \Big( \bpsi^\s \gamma_{[\s} 
\gamma^{\m\n\rh} \gamma_{\tau]} \psi^\tau +4\bpsi_\s \gamma^{\s\m\n\rh}\chi 
- 4{\bar\chi}\gamma^{\m\n\rh}\chi\Big)\, \biggr]\ ,
\label{BdR1}
\w2
\mathcal{L}_{\alpha'} (R^2) 
&= \alpha' e e^{2\vp} \Bigl[ 
- \tfrac14 R_{\m\n rs}(\Omega_-) R^{\m\n rs}(\Omega_-) 
-2R_{\m\n rs}(\Omega_-)  
\bpsi^r \gamma^\n D^\m (\omega,\Omega_-)\psi^s
%%%%%
\nn
\w2
&  +\tfrac12 R_{\m\n}{}^{rs}(\Omega_-)  
\bpsi_{rs} \left(\gamma^\rh \gamma^{\m\n} \psi_\rh + 2\gamma^{\m\n}\chi \right)
- \bpsi^{rs}\gamma^\m D_\m(\omega,\Omega_-)\psi_{rs}
\nn\w2
& - \tfrac{1}{12} H_{\m\n\rh} \bpsi^{rs} \gamma^{\m\n\rh} \psi_{rs} \Bigr]\ ,
\label{BdR2}
\end{align}}
where $\Gamma = \big\{ \big\}$ refers to the Christoffel symbol, and 
\be
\Omega_{\pm \m rs} = \omega_{\m rs}\pm H_{\m rs}\ ,\qquad H_{\m\n\rh} = 3\partial_{[\m} B_{\n\rh]}\ . 
\ee
The spin connection $\omega_{\m rs}$ is the standard torsion-free one, following from $D_\m (\omega,\Gamma) e_\n{}^r=0$, sometimes denoted by $\omega_{\m rs}(e)$. On the other hand, $\omega^L_{\mu\nu\rho}$ is the Lorentz Chern-Simons form
\be
\omega^L_{\m\n\rh}(\Omega_-) = \tr \left(
\Omega_{-[\m} \partial_\n \Omega_{-\rh]} + \tfrac23 \Omega_{-[\m}\Omega_{-\n} 
\Omega_{-\rh]} \right)\ .
\ee
The Lagrangian $\mathcal{L}_{0, \, \mathcal{O}(\alpha)}$ can be absorbed to the $H$-dependent terms in $\cL_0$ by letting 
\be
H_{\mu\nu\rho} = 3\partial_{[\m} B_{\n\rh]} 
\ \to \ \cH_{\m\n\rh}=3\partial_{[\m} B_{\n\rh]} 
-6\alpha' \omega^L_{\m\n\rh}(\Omega^{(sc)}_-)\ ,
\label{cH}
\ee
where $\Omega^{(sc)}_{-\mu rs}$ is the supercovariantized $\Omega_{-\mu rs}$
given as\footnote{Supercovariantized objects are usually denoted by hatted symbols. Here we use the unusual notation (sc) to indicate supercovariantizations instead, because we save the hatting for the $10D$ objects, when we consider the dimensional reduction.}
\bea
\Omega^{(sc)}_{-\m rs} &=& \Omega_{-\m rs} + T_{\m rs} \ ,
\nn\w2
T_{\m rs} &=& \big(\bpsi_\m \gamma_{[r}\psi_{s]} +\tfrac12\bpsi_r \gamma_\m \psi_s \big) -\tfrac32\bpsi_{[\m}\gamma_r\psi_{s]} = \bpsi_r \gamma_\m \psi_s\ .
\label{OMsc}
\eea
The second term in $\cL_{\alpha'}(R^2)$ arises  from ${\rm Riem}^2$ term 
through the fermionic torsion dependence in $\Omega^{(sc)}_{-\mu rs}$. 
Further definitions are 
\begin{align}
\psi_{rs} 
&= 2 e_r{}^\m e_s{}^\n
D_{[\mu}(\Omega_+)\psi_{\n]}\ , 
%%%%%%%%
\label{gcurv}\w2
D_\m (\omega,\Omega_-)\psi_{rs}
&= \left( \partial_\m 
+ \tfrac14 \omega_{\m pq} \gamma^{pq} \right) \psi_{rs} 
+\Omega_{-\m r}{}^p \psi_{ps} + \Omega_{-\m s}{}^p\psi_{r p}\ . 
\label{dgcurv}
\end{align}
The covariant derivative $D_\m (\omega,\Omega_-)\psi_{rs}$ requires extreme care, due to its unusual form in which the spinor indices are rotated by torsion free connection $\omega$, while the Lorentz vector indices are rotated by the torsionful connection $\Omega_-$. This asymmetric occurrence of the spin connection arises because the construction of $\cL_{\alpha'}(R^2)$ relies on treating $R_{\m\n rs}(\Omega_-^{(sc)})$ as Lorentz algebra valued Yang-Mills curvature \cite{Bergshoeff:1986vy,Bergshoeff:1989de}. 

The action of the Lagrangian \eq{Lag} is invariant under the following supersymmetry transformation rules up to ${\cal O}(\alpha'^2)$, and cubic fermion terms,
\begin{align}
\delta e_\m{}^r
&= {\bar\e} \gamma^r \psi_\m\ , \nonu 
\delta\psi_\m
&= D_\m(\Omega_+) \epsilon  -\tfrac32 \alpha'\, \omega^L_{\m\n\rh} \gamma^{\n\rh} \epsilon \ , \nonu
\delta B_{\m\n}
&= - {\bar\e} \gamma_{[\m} \psi_{\nu]} + 2 \alpha' \,\big( \Omega_{-[\mu}{}^{rs}\delta\Omega^{(sc)}_{-\nu] rs}\big) \ , \nonu 
\delta \chi
&= \frac{1}{2} \gamma^\m \e \partial_\m \vp -\frac{1}{12} H_{\m\n\rh}\gamma^{\m\n\rh}\e +\frac12\alpha'\, \omega^L_{\m\n\rh} \gamma^{\m\n\rh}\e\ , \nonu
\delta \vp &= {\bar\e}\chi\ .
\label{10dsuper}
\end{align}
The $\alpha'$ dependent terms in $\delta\psi_\m$ and $\delta\chi$ can be absorbed into to the definition of $H$ by letting $H\to \cH$ as in \eq{cH}, but we will work with $H=dB$ and exhibit the $\alpha'$ dependent terms explicitly, as we have been doing so far. Furthermore, $\Omega^{(sc)}_{-\m rs}$ defined in \eq{OMsc} transforms under supersymmetry as
\be
\delta  \Omega^{(sc)}_{-\m rs} =  -{\bar\e} \gamma_\m \psi_{rs}\ ,
\ee
%

%%%%%%%%%%%%%%%%%%%%%%%%%%%%%%%%%%%%%%%%%%%%%%%%%%%%%%%%%%%%%%%%%%%%%%
\section{The set up for dimensional reduction}
%%%%%%%%%%%%%%%%%%%%%%%%%%%%%%%%%%%%%%%%%%%%%%%%%%%%%%%%%%%%%%%%%%%%%%

We shall study the ordinary dimensional reduction on $T^4$. From here on, we put hats on all the fields and indices of $10D$ fields, and decompose the indices as $\hat\m = (\m, \alpha)$ and $\hat r= (r, a)$ where $\m, r=0,1,...,5$ and $\alpha, a =1,...,4$. For further notation and conventions, see Appendix A. As we truncate supersymmetry from $(1,1)$ to $(1,0)$, we take the $10D$ vielbein to be
\begin{equation}
\hat{e}_{\hat\mu}{}^{\hat r} 
= \left( 
\begin{array}{cc}
e_\mu{}^r & 0 \\
0 & E_\alpha{}^a
\end{array}
\right),  
\end{equation}
where off-diagonal vector components have been set to zero. As a result, the nonvanishing components of ${\hat\omega}_{\hat\m \hat r \hat s}$ are
\begin{align}
\hat{\omega}_{\mu rs} &= \omega_{\mu rs}\ , 
\qquad 
\hat{\omega}_{\mu ab} = \QZ_{\mu ab}\ , 
\qquad
\hat{\omega}_{\alpha ra} = - E_\alpha{}^b \PZ_{rab}\ ,
\end{align}
where
\begin{equation}
\QZ_{\mu ab} := E_{[a|}{}^\alpha \partial_\mu E_{\alpha |b]}\ ,
\qquad
\PZ_{\mu ab} := E_{(a|}{}^\alpha \partial_\mu E_{\alpha |b)}\ .
\label{qpzero}
\end{equation}
The nonvanishing Riemann tensor components are
\begin{align}
\hat{R}_{\mu\nu rs} ({\hat\omega})&= R_{\mu\nu rs} (\omega)\ , 
\nn\w2
\hat{R}_{\mu\nu ab} ({\hat\omega}) &= \QZ_{\mu\nu ab}\ , 
\nn\w2
\hat R_{\m a \n b} =& - D_\m(\Gamma) \PZ_{\n ab} - {\widetilde X}_{\m\n ab} \ ,
\nn\w2
\hat{R}_{ab rs}({\hat\omega}) &= \QZ_{rs ab}\ ,
\nn\w2
\hat R_{ab}{}^{cd} =& -2 \PZ_{\m[a}{}^c \PZ^\m{}_{b]}{}^d\ ,
\end{align}
where
\begin{align}
\QZ_{\mu\nu ab} &:= \partial_\mu \QZ_{\nu ab} + \QZ_{\mu a}{}^c \QZ_{\nu cb} 
- (\mu \leftrightarrow \nu)\ ,
\nn\w2
{\widetilde X}_{\m\n ab} &= \PZ_{\m a}{}^c \PZ_{\n cb}\ ,
\nn\w2
D_\m (\Gamma) \PZ_{\n ab} &= \partial_\m \PZ_{\n ab} -\Gamma_{\m\n}{}^\rh \PZ_{\rh ab} 
+ \QZ_{\m a}{}^c \PZ_{\n cb} + \QZ_{\mu b}{}^c \PZ_{\n ac}\ .
\end{align}
The $10D$ scalar curvature is
\begin{equation}
\hat{R} = R - 2 D_\mu \PZ^{\mu a}{}_a - \PZ_{\mu ab} \PZ^{\mu ab} 
- \PZ^{\mu a}{}_a \PZ_{\mu}{}^b{}_b\ . 
\end{equation}
We also decompose the 2-form potential as
\begin{equation}
\hat{B}_{\hat{\mu}\hat{\nu}} 
= ( B_{\mu\nu}, \ B_{\mu\alpha}=0, \ B_{\alpha\beta})\ . 
\end{equation}
Its field strength $\hat{H}_{\hat{\mu}\hat{\nu}\hat{\rho}} 
= 3 \partial_{[\hat{\mu}} \hat{B}_{\hat{\nu}\hat{\rho}]}$ 
has the only non-vanishing components 
\bea
\hat{H}_{\mu\nu\rho} &=& H_{\mu\nu\rho} 
:= 3 \partial_{[\mu} B_{\nu\rho]}\ ,
\nn\w2
\hat{H}_{\mu\alpha\beta} &=& \partial_\mu B_{\alpha\beta}\ .
\eea
In order to uncover the parametrization of the coset 
\be
Gr(4,4) = \frac{SO(4,4) }{ SO(4)_+ \times SO(4)_- }\ ,
\ee
by scalar fields other than the dilaton, we introduce the $SO(4,4)$-valued field 
\begin{align}
V &= \left( 
\begin{array}{cc}
V_a{}^\alpha & V_a{}_\alpha \\
V^{a\alpha} & V^a{}_\alpha
\end{array}
\right)
= \left( 
\begin{array}{cc}
E_a{}^\alpha & -2E_a{}^\beta B_{\beta\alpha} \\
0 & E_\alpha{}^a
\end{array}
\right)\ ,
\end{align}
which satisfies 
$V^T \eta V = \eta$, where 
$\eta = \left( 
\begin{array}{cc}
0 & 1 \\
1 & 0 
\end{array}
\right)$. The Maurer-Cartan form is 
\be
V \partial_\mu V^{-1} = 
\left(
\begin{array}{c|c}
E_a{}^\alpha \partial_\mu E_\alpha{}^b \  & 
 \ 2 E_a{}^\alpha E_b{}^\beta \partial_\mu B_{\alpha\beta} \\
\hline
0 & -E_b{}^\alpha \partial_\mu E_\alpha{}^a
\end{array}
\right)\ .
\ee
Changing the basis $W=\rho^T V \rho$ where 
$\rho= \frac{1}{\sqrt 2}  \left(\begin{array}{cc} 1& -1 \\ 1 & 1 \end{array} \right)$, 
which diagonalizes $\eta$ as $\rho^T \eta \rho = \left( 
\begin{array}{cc}
 1 & 0 \\
0 & -1
\end{array}
\right)$, gives
\begin{equation}
W \partial_\mu W^{-1} = \left( 
\begin{array}{cc}
Q_{+ \mu ab} & - P_{- \mu ab} \\
- P_{+ \mu ab} & Q_{- \mu ab}
\end{array}
\right)\ , 
\end{equation}
where $Q_{\pm \mu ab}$ and $P_{\pm \mu ab}$ are defined by
\begin{align}
Q_{\pm \mu ab} &:= E_{[a|}{}^\alpha \partial_\mu E_{\alpha |b]} 
\pm E_a{}^\alpha E_b{}^\beta \partial_\mu B_{\alpha\beta}\ , 
\nn\w2
P_{\pm \mu ab} &:= E_{(a|}{}^\alpha \partial_\mu E_{\alpha |b)} 
\pm E_a{}^\alpha E_b{}^\beta \partial_\mu B_{\alpha\beta}\ , 
\label{qp}
\end{align}
and they satisfy 
\begin{equation}
Q_{\pm \mu ab} = - Q_{\pm \mu ba}\ , \qquad
P_{+ \mu ab} = P_{- \mu ba}\ . 
\end{equation}
Thus, $Q_{\pm \m ab}$ are the composite connections associated with $SO(4)_\pm$. Note that they are related to each other as
\be 
Q_{+\m ab} -Q_{-\m ab} = P_{+\m ab} - P_{-\m ab}\ .
\ee
The equations \eq{qp} play central role in uncovering the $SO(4,4)$ symmetry of the dimensionally reduced action, through the use of the relations they imply such as 
\be
E_a{}^\alpha \partial_\mu E_{\alpha b} = Q_{+\m ab} +P_{-\m ab}\ ,\qquad 
2 E_a{}^\alpha E_b{}^\beta \partial_\mu B_{\alpha\beta} = P_{+\m ab}- P_{-\m ab}\ .
\ee
Other key relations follow from the Maurer-Cartan equation 
$ d (WdW^{-1}) + WdW^{-1} \wedge WdW^{-1} = 0$, which gives
\begin{align}
Q_{+\mu\nu ab} &= -2 P_{-[\mu|a}{}^c P_{+|\nu]cb}\ , 
\nn\w2
Q_{-\mu\nu ab} &= -2 P_{+[\mu|a}{}^c P_{-|\nu]cb}\ , 
\nn\w2
\partial_{[\mu|} P_{-|\nu]ab} &+ Q_{+[\mu|a}{}^c P_{-|\nu]cb} 
+ Q_{-[\mu|b}{}^c P_{-|\nu]ac} = 0\ ,  
\nn\w2
\partial_{[\mu|} P_{+|\nu]ab} &+ Q_{-[\mu|a}{}^c P_{+|\nu]cb} 
+ Q_{+[\mu|b}{}^c P_{+|\nu]ac} = 0\ ,    
\end{align}
where 
\begin{align}
Q_{+\mu\nu ab} &:= 2 \partial_{[\mu|} Q_{+ |\nu]ab} 
+ 2 Q_{+[\mu| a}{}^c Q_{+|\nu]cb}\ , 
\nn\w2
Q_{-\mu\nu ab} &:= 2 \partial_{[\mu|} Q_{- |\nu]ab} 
+ 2 Q_{-[\mu| a}{}^c Q_{-|\nu]cb}\ .
\end{align}
Note also the identity 
\begin{align}
\partial_\mu E_\alpha{}^a + Q_{\pm\mu}{}^a{}_b E_\alpha{}^b &= P_{\pm\mu}{}^a{}_b E_\alpha{}^b\ .
\label{pr}
\end{align}
In the rest of the paper, we shall use the notation
\be
P_{-\m ab} := P_{\m ab}\ , \quad P_{+\m ab} = P_{\m ba}\ .
\label{PT}
\ee
It is important to note $P_{\m ab}$ transforms under $SO(4)_\pm$ as
\be
\delta P_{\m ab} = \Lambda_{+a}{}^c P_{\m\,cb} + \Lambda_{-b}{}^c P_{\m\,ac}\ .
\label{gt}
\ee
Turning to the fermionic fields, we write SO(1,9) gamma matrices  $\hat{\gamma}^{\hat{m}}$ as 
\begin{align}
\hat{\gamma}^m &= \gamma^m \otimes 1 \quad (m=0,1,\cdots,5)\ , 
\nn\w2
\hat{\gamma}^{a+5} &= \gamma_7 \otimes \gamma^a \quad (a=1,2,3,4)\ ,
\label{10dgamma}
\end{align}
where $\gamma^m$ and $\gamma^a$ are SO(1,5) and SO(4) gamma matrices 
respectively, and $\gamma_7$ is the SO(1,5) chirality matrix. The SO(1,9) chirality matrix is 
\begin{equation}
\hat{\gamma}_{11} = \gamma_7 \otimes \gamma_5\ , 
\end{equation}
where $\gamma_5$ is the SO(4) chirality matrix. 
SO(4) gamma matrices and chirality matrix can be represented as 
\begin{align}
\gamma^a &= \left( 
\begin{array}{cc}
0 & i(\sigma^a)_{AB'} \\
-i(\bar{\sigma}^a)^{A'B} & 0
\end{array}
\right) \qquad (A,B=1,2,\ A',B'=1,2)\ , 
\nn\w2
\gamma_5 &= \gamma^1 \gamma^2 \gamma^3 \gamma^4 
= \left( 
\begin{array}{cc}
\delta_A{}^B & 0 \\
0 & - \delta^{A'}{}_{B'}
\end{array}
\right)\ , 
\end{align}
where 
\begin{equation}
\sigma^a = (\sigma^1, \sigma^2, \sigma^3, i)\ ,  \quad
\bar{\sigma}^a = (\sigma^1, \sigma^2, \sigma^3, -i)\ . 
\end{equation}
A general $10D$ spinor has components 
\begin{equation}
\hat{\psi} = \left( 
\begin{array}{c}
\psi_A \\
\psi^{A'}
\end{array}
\right)\ .
\end{equation}
In dimensional reduction we truncate the spinor fields as 
\begin{equation}
\hat{\psi}_\mu = \left( 
\begin{array}{c}
\hat{\psi}_{\mu A} \\
0
\end{array}
\right), \quad
\hat{\psi}_\alpha = \left( 
\begin{array}{c}
0 \\
\hat{\psi}_{\alpha}^{A'}
\end{array}
\right), \quad
\hat{\chi} = \left( 
\begin{array}{c}
\hat{\chi}_A \\
0
\end{array}
\right)\ , \quad
\hat{\epsilon} = \left( 
\begin{array}{c}
\hat{\epsilon}_A \\
0
\end{array}
\right)\ . 
\end{equation}
The $10D$ chirality conditions imply the following $6D$ chiralities
\begin{equation}
\gamma_7 \hat{\psi}_{\mu A} = + \hat{\psi}_{\mu A}\ , \quad
\gamma_7 \hat{\psi}_\alpha^{A'} = - \hat{\psi}_\alpha^{A'}\ , \quad
\gamma_7 \hat{\chi}_A = - \chi_A\ , \quad
\gamma_7 \hat{\epsilon}_A = + \hat{\epsilon}_A\ . 
\end{equation}

The $6D$ spinor fields are defined as 
\begin{equation}
\psi_{\mu A} = \hat{\psi}_{\mu A}\ , \quad
\psi_a{}^{A'} = E_a{}^\alpha \hat{\psi}_\alpha{}^{A'}\ , \quad
\chi_A = \hat{\chi}_A - \tfrac{1}{2} (\sigma^a)_{AB'} \psi_a{}^{B'}\ , \quad
\epsilon_A = \hat{\epsilon}_A\ .
\end{equation}
In what follows, we will use the notation 
\be
\Gamma^a := \gamma_7 \otimes \gamma^a\ ,\qquad \{\Gamma^a,\gamma^\m \}=0\ .
\ee
The indices $A$, $A'$ are raised and lowered as 
\begin{equation}
\psi^A = \epsilon^{AB} \psi_B\ , \quad
\psi_A = \psi^B \epsilon_{BA}\ , \quad
\epsilon^{AB} = \epsilon_{AB} = \left( 
\begin{array}{cc}
0 & 1 \\
-1 & 0
\end{array}
\right)
\end{equation}
and similar equations with primed indices $A', B'$. 
The $10D$ Dirac conjugate is 
\begin{equation}
\bar{\hat{\psi}} = \hat{\psi}^\dagger i \hat{\gamma}^0  
= \left( (\psi_A)^\dagger i \gamma^0,\ (\psi^{A'})^\dagger i \gamma^0 \right) 
= \left( \bar{\psi}^A,\ - \bar{\psi}_{A'} \right)\ ,
\label{10ddiracconj}
\end{equation}
where $6D$ Dirac conjugates are defined as 
\begin{equation}
\bar{\psi}^A = (\psi_A)^\dagger i \gamma^0\ , \qquad
\bar{\psi}^{A'} = (\psi_{A'})^\dagger i \gamma^0\ .
\end{equation}
The $10D$ Majorana condition is 
\begin{equation}
\hat{\psi} = C_{10} \bar{\hat{\psi}}^T\ , 
\label{10dmajorana}
\end{equation}
where $C_{10}$ is an $SO(1,9)$ charge conjugation matrix satisfying 
\begin{equation}
C_{10}^{-1} \hat{\gamma}^{\hat{m}} C_{10} 
= - \hat{\gamma}^{\hat{m}T}\ , 
\qquad C_{10}^T = - C_{10}\ . 
\end{equation}
For the representation of $\hat{\gamma}^{\hat{m}}$ in (\ref{10dgamma}) 
$C_{10}$ can be chosen as 
\begin{equation}
C_{10} = C_6 \otimes C_{4}\ , 
\end{equation}
where $C_{6}$ and $C_{4}$ are $SO(1,5)$ and $SO(4)$ charge conjugation 
matrices respectively satisfying 
\begin{align}
C_{6}^{-1} \gamma^m C_{6} &= - \gamma^{mT}\ , 
\qquad C_{6}^T = C_{6}\ , 
\nn\w2
C_{4}^{-1} \gamma^a C_{4} &= \gamma^{aT}, 
\qquad C_{4}^T = - C_{4}\ .
\end{align}
The explicit form of $C_4$ is 
\begin{equation}
C_{4} = \left( 
\begin{array}{cc}
-\epsilon_{AB} & 0 \\
0 & - \epsilon^{A'B'}
\end{array}
\right)\ , \qquad
C_{4}^{-1} = \left( 
\begin{array}{cc}
\epsilon^{AB} & 0 \\
0 & \epsilon_{A'B'}
\end{array}
\right)\ .
\end{equation}
The $10D$ Majorana condition (\ref{10dmajorana}) on (\ref{10dgamma}) 
implies $6D$ symplectic Majorana conditions 
\begin{equation}
\psi^A = \epsilon^{AB} C_{6} \bar{\psi}_B^T\ , \qquad
\psi^{A'} = \epsilon^{A'B'} C_{6} \bar{\psi}_{B'}^T\ . 
\end{equation}
In this notation, we have, for example,
\be
\Gamma^a \psi^b = -\sigma^a \psi^b\ ,\qquad \Gamma^a \epsilon = {\bar\s}^a \epsilon\ .
\ee
Note also the `flipping' property
\be
\bpsi_1 \Gamma^{a_1...a_n} \gamma^{\m_1...\m_m} \psi_2 = (-1)^{n+m} \bpsi_2 \gamma^{\m_m...\m_1} \Gamma^{a_n...a_1} \psi_1\ ,
\ee
where $\psi_1$ and $\psi_2$ are any two symplectic Majorana-Weyl spinors in $6D$.

%%%%%%%%%%%%%%%%%%%%%%%%%%%%%%%%%%%%%%%%%%%%%%%%%%%%%%%%%%%%%%%%%%%%%%
\section{Dimensional reduction of $\cL_0$ }
%%%%%%%%%%%%%%%%%%%%%%%%%%%%%%%%%%%%%%%%%%%%%%%%%%%%%%%%%%%%%%%%%%%%%%

From the $10D$ supertransformations (\ref{10dsuper})  we obtain the $6D$ supertransformations at zeroth order in $\alpha'$ as 
\begin{align}
\delta_0 e_\mu{}^m &= \bar{\epsilon} \gamma^m \psi_\mu\ , \nonu 
\delta_0 \psi_\mu &= D_\mu(\Omega_+) \epsilon\ , \nonu
\delta_0 B_{\mu\nu} 
&= - \bar{\epsilon} \gamma_{[\mu} \psi_{\nu]}\ , \nonu
\delta_0 \chi 
&= \tfrac{1}{2} \gamma^\mu \epsilon \partial_\mu  \varphi 
- \tfrac{1}{12} H_{\mu\nu\rho} \gamma^{\mu\nu\rho} \epsilon\ , 
\nonu 
\delta_0 \varphi &= \bar{\epsilon} \chi\ , \nonu 
W \delta_0 W^{-1} 
&=  \left( 
\begin{array}{cc}
0 & - \bar{\epsilon} \Gamma_a \psi_b \\
- \bar{\epsilon} \Gamma_b \psi_a & 0
\end{array}
\right), 
\nonu
\delta_0 \psi_a 
&= - \tfrac12 \gamma^\mu \Gamma^b \epsilon P_{\mu ba} \ ,
\label{susy1}
\end{align}
where we have defined the $6D$ dilaton $\varphi$ as 
\begin{equation}
\varphi = \hat{\varphi} + \tfrac12 \ln E\ , 
\qquad E = \det E_\alpha{}^a\ ,
\end{equation}
and the covariant derivative on $\epsilon$ is given by 
\begin{equation}
D_{\mu}(\Omega_+) \epsilon
= \left( \partial_{\mu} + \tfrac{1}{4} \Omega_{+\mu mn} \gamma^{mn} 
+ \tfrac{1}{4} Q_{+\mu ab} \Gamma^{ab} \right) \epsilon\ .
\end{equation}
We have suppresses the connection $Q_+$ in the covariant derivative $D_\mu(\Omega_+)\epsilon$, in accordance with our notational convention described in Appendix A.
The supertransformations of the hyperscalars are obtained by using
\be
\delta_0 E_\alpha{}^a = \bar{\epsilon} \Gamma^a \psi_\alpha\ ,\qquad  \delta_0 B_{\alpha\beta}
= -\bar{\epsilon} \Gamma_{[\alpha} \psi_{\beta]}\ .
\ee
To begin with this gives
\begin{equation}
W \delta_0 W^{-1} 
=\left(
\begin{array}{c|c}
- 2 \bar{\epsilon} \Gamma_{[a} \psi_{b]} & - \bar{\epsilon} \Gamma_a \psi_b \\
\hline
- \bar{\epsilon} \Gamma_b \psi_a & 0
\end{array}
\right)\ .
\end{equation}
We can add a compensating $SO(4)_+$ transformation 
$\delta_{SO(4)_+}$ such that $W \delta W^{-1}$ 
takes values only in the coset direction: 
\begin{align}
W (\delta_0 + \delta_{SO(4)_+} ) W^{-1} 
&= \left( 
\begin{array}{c|c}
- 2 \bar{\epsilon} \Gamma_{[a} \psi_{b]} - \lambda_{+ab} 
& - \bar{\epsilon} \Gamma_a \psi_b \\
\hline
- \bar{\epsilon} \Gamma_b \psi_a & 0 
\end{array}
\right)
\nn\w2
&= \left( 
\begin{array}{cc}
0 & - \bar{\epsilon} \Gamma_a \psi_b \\
- \bar{\epsilon} \Gamma_b \psi_a & 0
\end{array}
\right), 
\label{dW2}
\end{align}
where we have chosen the $SO(4)_+$ transformation parameter  as $\lambda_{+ab} = - 2 \bar{\epsilon} \Gamma_{[a} \psi_{b]}$.  In \eq{susy1}, we have denoted this result as $W\delta_0 W^{-1}$ for short. Other fields which transform under $SO(4)_+$ are fermi fields,  for which this compensating transformation is higher order in fermi fields and can be ignored. $B_{\mu\nu}$ transforms only under $SO(4)_-$. For later convenience, let us also record the supertransformations
\begin{align}
\delta_0 Q_{-\mu ab} &= 2 P_{\mu c[a} \bar{\epsilon} \Gamma^c \psi_{b]}\ , 
\nn\w2
\delta_0 Q_{+\mu ab} &= 2 P_{\mu c[a} \bar{\epsilon} \Gamma_{|b]} \psi^c
+ 2 D_\mu(Q_+, Q_-) (\bar{\epsilon} \Gamma_{[b} \psi_{a]})
\nn\w2
&= 2 P_{\mu [a}{}^c \bar{\epsilon} \Gamma_{b]} \psi_c
+ 2 D_\mu(Q_+, Q_+) (\bar{\epsilon} \Gamma_{[b} \psi_{a]}), 
\nn\w2
\delta_0 P_{\mu ab} &= D_\mu(Q_+,Q_-) (\bar{\epsilon}\Gamma_a \psi_b) 
+ 2 \bar{\epsilon} \Gamma_{[a} \psi_{c]} P_\mu{}^c{}_b\ . 
\end{align}
The covariant derivatives are defined as
\begin{align}
D_\m(Q_+, Q_-) ( \bar{\e} \Gamma_b \psi_a ) =&\, \pd_\m ( \bar{\e} \Gamma_b \psi_a ) + Q_{+\m b}{}^c ( \bar{\e} \Gamma_c \psi_a ) + Q_{-\m a}{}^c ( \bar{\e} \Gamma_b \psi_c ) \ , 
\nn\w2
D_\m(Q_+, Q_+) ( \bar{\e} \Gamma_b \psi_a ) =&\, \pd_\m ( \bar{\e} \Gamma_b \psi_a ) + Q_{+\m b}{}^c ( \bar{\e} \Gamma_c \psi_a ) + Q_{+\m a}{}^c ( \bar{\e} \Gamma_b \psi_c ) \ . 
\end{align}

$\delta_0 Q_{-\mu ab}$ has the right $SO(4)_+ \times SO(4)_-$ index structure.
$\delta_0 Q_{+\mu ab}$ has undesirable index structures in the first line. 
But it can be written as in the second line, in which the first term has the right index structure 
and the second term is a local $SO(4)_+$ transformation.  So, if we add a compensating $SO(4)_+$ 
transformation with the same parameter $\lambda_{+ab}$ as in \eq{dW2} to the supertransformation 
so that the second term is cancelled, we obtain the right index structure.  The second term of 
$\delta_0 P_{\mu ab}$ has a undesirable index structure but it is also a local $SO(4)_+$ transformation 
with the same parameter  as for $\delta_0 Q_{+\mu ab}$. 
Supertransformations of the truncated components automatically vanish
\begin{equation}
\delta_0 \ ( \ \hat{e}_\mu{}^a\ ,\  
\hat{e}_\alpha{}^m\ , \ 
\hat{B}_{\mu\alpha}\ , \ 
\hat{\psi}_{\mu}{}^{A'}\ , \ 
\hat{\psi}_{\alpha A}\ , \ 
\hat{\chi}_{\alpha}{}^{A'} \ ) = 0\ ,
\end{equation}
which shows the consistency of the truncation from $(1,1)$ to $(1,0)$ supersymmetry. 

Using the ingredients described in considerable detail above, it is now straightforward to perform the dimensional 
reduction of the two-derivative Lagrangian $\cL_0$ given in \eq{het}, which yields the $6D$ Lagrangian
\begin{align}
\mathcal{L}_0 &= e e^{2\varphi} \Big[ \, \tfrac{1}{4} R 
+ g^{\mu\nu} \partial_\mu \varphi \partial_\nu \varphi 
- \tfrac{1}{12} H_{\mu\nu\rho} H^{\mu\nu\rho} - \tfrac{1}{4} P_{\mu ab} P^{\mu ab}
\nn\\ &\quad 
- \tfrac{1}{2} \bar{\psi}_\mu \gamma^{\mu\nu\rho} 
D_\nu(\omega) \psi_\rho 
+ 2 \bar{\chi} \gamma^{\mu\nu} D_\mu(\omega) \psi_\nu 
+ 2 \bar{\chi} \gamma^\mu D_\mu(\omega) \chi 
\nn\\&\quad
 - \tfrac{1}{2} \bar{\psi}^a \gamma^\mu D_\mu(\omega) \psi_a  - \partial_\mu \varphi \left( 
\bar{\psi}^\mu \gamma^\nu \psi_\nu 
 + 2 \bar{\psi}_\nu \gamma^\mu \gamma^\nu \chi \right) 
\nn\\&\quad
+ \tfrac{1}{2} P_{\mu ab} \Big( 
\bar{\psi}_\nu \gamma^\mu \gamma^\nu \Gamma^a \psi^b 
+ 2 \bar{\chi} \gamma^\mu \Gamma^a \psi^b \Big) 
- \tfrac{1}{24} H_{\mu\nu\rho} \Big( 
\bar{\psi}^\sigma \gamma_{[\sigma} \gamma^{\mu\nu\rho} 
\gamma_{\tau]} \psi^\tau 
\nn\\&\quad 
+ 4 \bar{\psi}_\sigma \gamma^{\sigma\mu\nu\rho} \chi 
- 4 \bar{\chi} \gamma^{\mu\nu\rho} \chi +  \bar{\psi}^a \gamma^{\mu\nu\rho} \psi_a   \Big) 
\Big]\ .
\label{6D}
\end{align}
The definitions of the covariant derivatives occurring above are listed in Appendix A.
In obtaining the $6D$ Lagrangian, we have also used the relations
\begin{align}
& D_\mu(\omega,\QZ) \psi_\alpha 
= E_\alpha{}^a D_\mu(\omega, \QZ) \psi_a 
+ \PZ_\mu{}^a{}_b E_\alpha{}^b \psi_a\ , 
\nn\w2
& \PZ_\mu{}^a{}_a = E^{-1} \partial_\mu E, 
\qquad
- \PZ_{\mu ab} \PZ^{\mu ab} = \tfrac14 \partial_\mu G^{\alpha\beta} 
\partial^\mu G_{\alpha\beta}\ , 
\nn\w2
& G_{\al\beta}:= E_\al{}^a E_{\beta a}\ ,\qquad 
P_{\pm \mu ab} =  \PZ_{\mu ab} 
\pm E_a{}^\alpha E_b{}^\beta \partial_\mu B_{\alpha\beta}\ .
\end{align}
We conclude by emphasizing that the $6D$ Lagrangian \eq{6D} and supertransformations \eq{susy1} are manifestly $SO(4,4)$ invariant, 
as well as $SO(4)_+\times SO(4)_-$ symmetric. The lowest order bosonic field equations are
\begin{align}
\mathcal{E}_\vp =&\, \tfrac12 R(\om) - 2\, \square \vp - 2 ( \pd_\m \vp ) ( \pd^\m \vp) - \tfrac16 H_{\m\n\rh} H^{\m\n\rh} - \tfrac12 P_{\m a b} P^{\m a b} \ , 
\nn\w4
\mathcal{E}_{\m\n} =&\, \tfrac14 R_{\m\n}(\om) - \tfrac12 \left( D_\m(\Gamma) \pd_\n \vp \right)
 - \tfrac14 H_{\m\rh\s} H_\n{}^{\rh\s} - \tfrac14 P_{\m a b} P_\n{}^{ab} - \tfrac14 \mathcal{E}_\vp g_{\m\n} \ , 
\nn\w4
\mathcal{B}_{\m\n} =&\, D^\rh(\Gamma) ( e^{2 \vp} H_{\m\n\rh} ) \ , 
\nn\w4
\mathcal{E}_P^{ab} =&\, D_\m ( e^{2 \vp} P^{\m a b} ) \ , 
\label{bosfe}
\end{align}
where $\mathcal{E}_\vp$, $\mathcal{E}_{\m\n}$, $\mathcal{B}_{\m\n}$, and $\mathcal{E}_P^{ab}$ are dilaton field equation, 
Einstein equation, $B$-field equation, and hyperscalar field equation respectively. The lowest order fermionic field equations are
\begin{align}
\mathcal{E}_\chi =&\, \g^\m D_\m(\om) \chi + \tfrac12 \g^{\m\n} D_\m(\om) \psi_\n + \tfrac14 \g^\m \Gamma^a \psi^b P_{\m a b} + \g^\m \chi \pd_\m \vp - \tfrac12 \g^\m \g^\n \psi_\m \pd_\n \vp 
\nn\w2
&\, + \tfrac1{24} \g^{\m\n\rh\s} \psi_\s H_{\m\n\rh} + \tfrac1{12} \g^{\m\n\rh} \chi H_{\m\n\rh} \ , 
\nn\w4
\mathcal{E}_\psi^\m =&\, \g^{\m\n\rh} D_\n(\om) \psi_\rh + 2 \g^{\m\n} D_\n(\om) \chi - \tfrac12 \g^\n \g^\m \Gamma^a \psi^b P_{\n a b} - \g^{\m\n\rh} \psi_\n \pd_\rh \vp + 4 \g^{\m\n} \chi \pd_\n \vp 
\nn\w2
&\, + \g^\n \psi_\n \pd^\m \vp - \g^\m \psi^\n \pd_\n \vp + 2 \g^\n \g^\m \chi \pd_\n \vp + \tfrac1{12} \g^{[\m} \g_{\rh\s\ta} \g^{\n]} \psi_\n H^{\rh\s\ta} + \tfrac16 \g^{\m\n\rh\s} \chi H_{\n\rh\s} \ , 
\nn\w4
\mathcal{E}_a =&\, \g^\m D_\m(\om) \psi_a + \tfrac12 \g^\m \g^\n \Gamma^b \psi_\m P_{\n b a} + \g^\m \Gamma^b \chi P_{\m b a} + \g^\m \psi_a \pd_\m \vp + \tfrac1{12} \g^{\m\n\rh} \psi_a H_{\m\n\rh} \ , 
\label{ferfe}
\end{align}
where $\mathcal{E}_\chi$, $\mathcal{E}_\psi^\m$, and $\mathcal{E}_a$ are $\chi$-field equation, gravitino field equation, and hyperino field equation respectively.

%%%%%%%%%%%%%%%%%%%%%%%%%%%%%%%%%%%%%%%%%%%%%%%%%%%%%%%%%%%%%%%%%%%%%%%%%%%%%%%%%%
\section{Dimensional reduction of $\OA$ terms }
%%%%%%%%%%%%%%%%%%%%%%%%%%%%%%%%%%%%%%%%%%%%%%%%%%%%%%%%%%%%%%%%%%%%%%%%%%%%%%%%%%%

\subsection{Building blocks}
%%%%%%%%%%%%%%%%%%%%%%%%%%%%%%%%%

We begin by the dimensional reduction of the $H$-torsionful Lorentz connection
\be
\hat{\Omega}_{\pm \hat{\mu}\hat{r}\hat{s}} 
= \hat{\omega}_{\hat{\mu}\hat{r}\hat{s}} 
\pm \hat{H}_{\hat{\mu}\hat{r}\hat{s}}\ ,
\label{Omega}
\ee
where $\hat\omega = \hat\omega(\hat e)$ and $\hat H=d\hat B$. Its dimensional reduction gives the only nonvanishing components 
\bea
\hat{\Omega}_{\pm \mu rs} &=& \omega_{\mu rs} \pm H_{\mu rs}\ , 
\nn\w2
\hat{\Omega}_{\pm \mu ab} &=& Q_{\pm \mu ab}\ , 
\nn\w2
\hat{\Omega}_{\pm \alpha ra} &=& - E_\alpha{}^b P_{\mp r ab}\ , 
\eea
where $\omega =\omega(e)$ and $H=dB$. It follows that the only nonvanishing components of the $10D$ Riemann tensor for $\hat\Omega_-$ are
\begin{align}
\hat{R}_{\mu\nu rs}(\hat{\Omega}_-) &= R_{\mu\nu rs}(\Omega_-)\ , 
\nn\w2
\hat{R}_{\mu\nu ab}(\hat{\Omega}_-) &= Q_{-\mu\nu ab}\ , 
\nn\w2
\hat R_{\m a \n b} =& - D_\m(\Gamma_+) P_{\n ab} - X_{\m\n ab} \ , 
\nn\w2
\hat{R}_{ab rs}(\hat{\Omega}_-) 
&= Q_{+rsab}\ ,
\nn\w2
\hat R_{ab}{}^{cd} =& -2 P_{\m[a}{}^c P^\m{}_{b]}{}^d\ , 
\label{riemann}
\end{align}
where 
\be
D_\m (\Gamma_+) P_{\n ab} = \partial_\m P_{\n ab} -\Gamma_{+\m\n}{}^\rh P_{\rh ab} 
+ Q_{+\m a}{}^c P_{\n cb} + Q_{-\mu b}{}^c P_{\n ac}\ ,
\label{DP}
\ee
and $\Gamma_{\pm \m\n}{}^\rh = \Gamma_{\m\n}{}^\rh \pm H_{\m\n}{}^\rh$. We shall often adhere to the convention in which the 
dependence of $Q_\pm$ in covariant derivatives will be suppressed if they act normally on the $SO(4)_+ \times SO(4)_-$ indices as above.

Next, we consider the dimensional reduction of the $10D$ Lorentz Chern-Simons form
\be
\hat{\omega}^L_{\hat{\mu}\hat{\nu}\hat{\rho}}
= \tr \left(
\hat{\Omega}_{-[\hat{\mu}} \partial_{\hat{\nu}} \hat{\Omega}_{-\hat{\rho}]}
+ \tfrac23 \hat{\Omega}_{-[\hat{\mu}} \hat{\Omega}_{-\hat{\nu}} 
\hat{\Omega}_{-\hat{\rho}]}
\right)\ .
\ee
Its only nonvanishing components are 
\begin{align}
\hat{\omega}^L_{\mu\nu\rho} &= \omega^L_{\mu\nu\rho}(\Omega_-) 
+\omega^Q_{\m\n\rh}(Q_-)\ , 
\nn\w2
\hat{\omega}^L_{\mu ab} &= \tfrac23 P_{\n a}{}^c \big( D_\m(\Gamma_+) P^\n{}_{bc} 
+ X_\m{}^\n{}_{bc} \big) \Big|_{[ab]}\ ,
\label{rcs}
\end{align}
where
\be
\omega^Q_{\m\n\rh}(Q_-) = \tr \left( Q_{-[\mu} \partial_\nu Q_{-\rho]} 
+ \tfrac23 Q_{-[\mu} Q_{-\nu} Q_{-\rho]} \right)\ .
\ee
We see that the Chern-Simons form built out of the composite local connection $Q_{-\m ab}$ naturally arises as a result of the dimensional reduction.

The building blocks for the fermionic Lagrangian at $\OA$ are as follows.  Components of ${\hat \psi}_{\hat{m}\hat{n}}$ defined in \eq{gcurv} decompose in $6D$ as
\begin{align}
\hat{\psi}_{rs} &= \psi_{rs} 
:= 2 D_{[r}(\Omega_+,Q_+) \psi_{s]}\ ,
\nn\w2
\hat{\psi}_{ra} &= E_a{}^\alpha D_r(\Omega_+, Q_+) \psi_\alpha 
+\tfrac12 P_{\mu ba} \gamma^\mu \Gamma^b \psi_r
\nn\w2
 &= D_r(\Omega_+, Q_+, Q_-) \psi_a + P_{r ba} \psi^b
+ \tfrac12 P_{\mu ba} \gamma^\mu \Gamma^b \psi_r\ ,
\nn\w2
\hat{\psi}_{ab} &= -P_{\mu c[a} \gamma^\mu \Gamma^c \psi_{b]}\ ,
\end{align}
and the components of ${\hat D}_{\hat\m}(\hat\omega, \hat\Omega_-) \hat{\psi}_{\hat r}$ in $6D$ are
\begin{align}
{\hat D}_\m(\hat\omega, \hat\Omega_-) \hat\psi_r =& D_\m (\omega,\Omega_-) \psi_r +\tfrac14 P_{\m ab} \Gamma^{ab}\psi_r
\nn\w2
{\hat D}_\m(\hat\omega, \hat\Omega_-) \hat\psi_a =& D_\m (\omega) \psi_a +\tfrac14 P_{\m cd} \Gamma^{cd}\psi_a\ ,
\nn\w2
{\hat D}_a(\hat\omega, \hat\Omega_-) \hat\psi_r =& -\tfrac12 P_{\m (ab)} \gamma^\m \Gamma^b \psi_r - P_{r ab} \psi^b \ ,
\nn\w2
{\hat D}_a(\hat\omega, \hat\Omega_-) \hat\psi_b =& -\tfrac12 \gamma^\m \Gamma^c P_{\m (ca)} \psi_b + P^\m{}_{ab} \psi_\m\ ,
\end{align}
where 
\be
D_\m (\omega,\Omega_-) \psi_r = \left(\partial_\m +\tfrac14 \omega_{\m pq} \gamma^{pq} +\tfrac14 Q_{+\m ab} \Gamma^{ab} \right)\psi_r + \Omega_{-\m r}{}^s \psi_s
\label{cd4}
\ee

\subsection{The bosonic Lagrangian at $\OA$ }
%%%%%%%%%%%%%%%%%%%%%%%%%%%%%%%%%%%%%%%%%%%%%%%%%%%%%%%%%%%%%%%

The first contribution to the bosonic Lagrangian at $\OA$ from \eq{BdR1} and \eq{BdR2} reduces as 
\begin{align}
& - \tfrac14 \hat{e} e^{2\hat{\varphi}} 
\hat{R}_{\hat{\mu}\hat{\nu}\hat{m}\hat{n}}(\hat{\Omega}_-)
\hat{R}^{\hat{\mu}\hat{\nu}\hat{m}\hat{n}}(\hat{\Omega}_-)
\nn\w2
& = e e^{2\varphi} \Bigl[ 
- \tfrac14 R_{\mu\nu mn}(\Omega_-) R^{\mu\nu mn}(\Omega_-) 
- \tfrac14 Q_{+\mu\nu ab} Q_+{}^{\mu\nu ab} - \tfrac14  Q_{-\mu\nu ab} Q_-{}^{\mu\nu ab} 
\nn\w2
&\quad -\Big( D_\mu(\Gamma_+) P_{\n ab} + X_{\m\n ab}\Big)
\Big( D^\mu(\Gamma_+) P^{\n ab} + X^{\m\n ab} \Big)
\nn\w2
&\quad -\tfrac12 Y_{\m\n} Y^{\m\n} +\tfrac12 Z_{\m\n ab} Z^{\m\n ba} \Bigr] \ ,
\label{LR2}
\end{align}
and the only other contribution to the bosonic Lagrangian at $\OA$ from \eq{BdR1} and \eq{BdR2} reduces as
\begin{align}
{\hat e} e^{2\hat \vp} \hat{H}^{\hat\m \hat\n \hat \rh} {\hat \omega}^L_{\hat\m \hat\n \hat \rh}
=&  e e^{2\vp} \Big[ H^{\m\n\rh} \big(\omega^L_{\m\n\rh} +\omega^Q_{\m\n\rh} \big)
+X^{\m\n ab} \Big( D_\m(\Gamma_+) P_{\n ab} + X_{\m\n ab} \Big) 
\nn\w2
& - Z^{\m\n ab} \Big( D_\m (\Gamma_+) P_{\n ab}  +X_{\m\n ab} \Big) \Big]\ .
\end{align}
Combining the two results we get
\begin{align}
\cL_B \Big|_\OA =&  {\hat e} e^{2\hat \vp} \Big[ \hat{H}^{\hat\m \hat\n \hat \rh} {\hat \omega}^L_{\hat\m \hat\n \hat \rh}
  - \tfrac14 \hat{R}_{\hat{\mu}\hat{\nu}\hat{m}\hat{n}}(\hat{\Omega}_-)
\hat{R}^{\hat{\mu}\hat{\nu}\hat{m}\hat{n}}(\hat{\Omega}_-) \Big]  
\nn\w2
= & e e^{2\varphi}\Bigg[ H^{\m\n\rh} \big( \omega^L_{\m\n\rh} + \omega^Q_{\m\n\rh} \big) 
- \tfrac14 R_{\mu\nu mn}(\Omega_-) R^{\mu\nu mn}(\Omega_-) - \tfrac14  Q_{+\mu\nu ab} Q_+{}^{\mu\nu ab}
\nn\w2
& \qquad - \tfrac14  Q_{-\mu\nu ab} Q_-{}^{\mu\nu ab}
-D_\mu(\Gamma_+) P_{\n ab}  D^\mu (\Gamma_+) P^{\n ab} 
-\tfrac12 Y_{\m\n} Y^{\m\n} +\tfrac12 Z_{\m\n ab} Z^{\m\n ba} 
\nn\w2
&  \qquad +\Delta\cL_B  \Bigg] \ ,
\label{h2r2}
\end{align}
where
\be
\Delta\cL_B  = -P^{\mu ab } D_\m Y_{ab}  - X^{ab} Y_{ab}\ .
\label{deltaLB}
\ee
Note that $Y_{ab}$ transforms as $SO(4)_+$ tensor, and the $Q_+$ connections acting on its indices have been suppressed. We have separated the terms called $\Delta\cL_B$ because they are the only terms in \eq{h2r2} that are not invariant  under $SO(4)_+\times SO(4)_-$. These terms break $SO(4)_+\times SO(4)_-$ down to the diagonal $SO(4)$. They will be removed by a field redefinition which will also produce a $SO(4)_+\times SO(4)_-$ invariant term $-e e^{2\vp} Z_{\m\n ab} Z^{\m\n ab}$, as will be shown in the next subsection. 

In obtaining \eq{h2r2} we have used the relations
\be
(X^{\m\n ab} + Z^{\m\n ab}) D_\m P_{\n ab} = P^{\m ab} D_\m Y_{ab}\ ,\qquad  Z^{\m\n ab} X_{\m\n ab} = X^{ab} Y_{ab}\ .
\ee
 We shall continue by reducing the fermionic Lagrangian at $\OA$ as well, and collecting all such problematic terms that break the $SO(4)_+\times SO(4)_-$ symmetry. The total of such terms will be called $\Delta\cL_F$ below. At the end we shall find the field redefinitions of the hyperscalar and hyperfermions which will remove $\Delta\cL_B +\Delta\cL_F$ completely, and producing few new terms that are $SO(4)_+\times SO(4)_-$ invariant.

\subsection{The fermionic Lagrangian at $\OA$ }
%%%%%%%%%%%%%%%%%%%%%%%%%%%%%%%%%%%%%%%%%%%%%%%%%%%%%%%%%%%%%%%

In addition to the definitions for the covariant derivatives given in \eq{DP} and \eq{cd4}, in what follows it is understood that
\begin{align}
D_\m \psi_a =& D_\m (\Omega_+)\psi_a = \left( \partial_{\mu} + \tfrac{1}{4} \Omega_{+\mu rs} \gamma^{rs} 
+ \tfrac{1}{4} Q_{+\mu cd} \Gamma^{cd} \right) \psi_a + Q_{-\mu a}{}^b \psi_b\ ,
\label{dpsi}
\w2
D_\m P_{\n ab} =& D_\m(\Gamma_+) P_{\n ab} =\partial_\m P_{\n ab} -\Gamma_{+\m\n}{}^\rh P_{\rh ab}  + Q_{+\m a}{}^c P_{\n cb} + Q_{-\mu b}{}^c P_{\n ac}\ ,
\label{DP2}
\end{align}
where $\Gamma_{+\m\n}{}^\rh  = \Gamma_{\m\n}{}^\rh  +H_{\m\n\rh}$ and $\Gamma_{\m\n}{}^\rh$ is the Christoffel symbol. Thus, we suppress the connections $Q_\pm$ in the covariant derivatives, if their action on the $SO(4)_+\times SO(4)_-$ indices is the standard one, as explained in Appendix A.

%\subsection{Desirable terms}
%%%%%%%%%%%%%%%%%%%%%%%%%%%%%%%%%

Next, we dimensionally reduce the fermionic Lagrangian at $\OA$ and in the order they appear in 
\eq{BdR1} and \eq{BdR2}. Not all terms that result from the dimensional reduction are $SO(4)_+\times SO(4)_-$ invariant. Such terms are invariant only under the diagonal subgroup, and they are collected as $\Delta\cL_i, i=1,...,7$ below. 
\allowdisplaybreaks{
\begin{align}
% old  \circled{10} + \circled{11} =  
\circled{1} =&\,  ee^{2\hat\vp} \Big[ -\hat H^{\hat\m\hat\n\hat\rh} \hat R_{\hat\m\hat\n}{}^{\hat r\hat s}({\hat\Omega}_{-}) \bpsi_{\hat r} \hat\gamma_{\hat \rh} \hat\psi_{\hat s}  + \hat\bpsi_{\hat r} \hat\gamma_{\hat \n}\hat\psi_{\hat s}\, \hat\Omega_{-\hat\rh}{}^{\hat r\hat s} \e^{-2\vp} D_{\hat\m}({\hat\Gamma}) \big(e^{2\vp}\hat H^{\hat\m\hat\n\hat\rh} \big) \Big] 
\nn\w2
= & ee^{2\vp} \Big[ -H^{\m\n\rh} R_{\m\n}{}^{rs}(\Omega_-)\,\bpsi_r \gamma_\rh \psi_s - H^{\m\n\rh} Q_{-\m\n}{}^{ab}\,\bpsi_a \gamma_\rh \psi_b
\nn\w2
&  +\big( \bpsi_r\gamma_\n\psi_s \Omega_{-\rh}{}^{rs} + \bpsi_a\gamma_\n\psi_b Q_{-\rh}{}^{ab}\big)  e^{-2\vp} D_\m (\Gamma)\big(e^{2\vp} H^{\m\n\rh}\big) +\Delta \cL_1 \Big]\ ,
%%%%%%%%%%%%%%%%%%%%%%%%%%%%%%%%%%%%%%%%%
\label{10plus11} \w4
% old \circled{2} = 
\circled{2} = &\, ee^{2\hat\vp} \Big[ \tfrac14 \hat{\omega}_{L\hat{\mu}\hat{\nu}\hat{\rho}}({\hat\Omega}_{-})
\Big( \bar{\hat{\psi}}^{\hat{\sigma}} \hat{\gamma}_{[\hat{\sigma}} 
\hat{\gamma}^{\hat{\mu}\hat{\nu}\hat{\rho}}  \hat{\gamma}_{\hat{\tau}]} \hat{\psi}^{\hat{\tau}} 
+4 \bar{\hat{\psi}}_{\hat{\sigma}}  \hat{\gamma}^{\hat{\sigma}\hat{\mu}\hat{\nu}\hat{\rho}} \hat{\chi} - 4 \bar{\hat{\chi}} \hat{\gamma}^{\hat{\mu}\hat{\nu}\hat{\rho}}\hat{\chi} \Big) \Big]
\nn\w2
= &\,  ee^{2\vp} 
\Big[  
\tfrac14 \big( \omega^L_{\m\n\rh}(\Omega_-) + \omega^Q_{\m\n\rh}(Q_-) \big)   
\Big( \bar{\psi}^\sigma \gamma_{[\sigma} \gamma^{\mu\nu\rho} 
\gamma_{\tau]} \psi^\tau  + 4 \bpsi_\sigma \gamma^{\sigma\mu\nu\rho} \chi 
\nn\w2
& - 4 \bar{\chi} \gamma^{\mu\nu\rho} \chi   +\bpsi^a \gamma^{\m\n\rh} \psi_a  \Big)  
+\Big( -\tfrac12\bpsi_\n \gamma^{\m\n\rh} \Gamma^{ab} \psi_\rh 
\nn\w2
& -2\bpsi_\n \gamma^{\m\n} \Gamma^{ab} \chi -2{\bar\chi} \gamma^\m \Gamma^{ab} \chi + \tfrac12\bpsi^d\gamma^\m \Gamma^{ab} \psi_d  \Big) P^\s{}_a{}^c D_\m P_{\s bc} +\Delta\cL_2 \Big] \ ,
%%%%%%%%%%%%%%%%%%%%%%%%%%%%%%%%%%%%%%%%%
\w4
% old \circled{9} =
\circled{3} =&\, ee^{2\hat\vp} \Big[ -2\hat{R}_{\hat{\mu}\hat{\nu}\hat{m}\hat{n}}(\hat{\Omega}_-)  
\bar{\hat\psi}^{\hat m} \hat\gamma^{\hat\n} D^{\hat\m} (\hat\omega,\hat\Omega_-) \hat\psi^{\hat n} \Big] 
\nn\w2
=&\, ee^{2\vp} \Big[-2 R^{\m\n rs}(\Omega_-) \bpsi_r \gamma_\n D_\m (\omega,\Omega_-)\psi_s 
-2 Q_{-\m\n}{}^{ab}  \bpsi_a \gamma^\n D^\m (\omega) \psi_b 
%%%%
\nn\w2
& +2\bpsi_r\Gamma_b\psi^c \Big(Q_+{}^{rs ab} P_{s ac}\Big) +2 \Big(\bpsi^r\Gamma^a D_\m (\omega)\psi^b -\bpsi^b\Gamma^a D_\m(\omega,\Omega_-)\psi^r\Big) D^\m P_{r ab}  
%%%%
\nn\w2
& -2\bpsi_r\gamma_\m\psi_\n \Big(P^\n{}_{ab} D_\m P_r{}^{ab}\Big) -2\bpsi_b\gamma_\m\psi_c \Big(P^{\n ac}  D_\m P_{\n a}{}^b\Big)  + 2 \bpsi^\n \Gamma^a \psi^b  P^\m{}_{ab} Y_{\m\n} 
\nn\w2
&  -2 \bpsi^\n \Gamma^a \psi^b \big(P^{\m c}{}_b Y_{\m\n ac}\big)  +2\big( \bpsi^\n\Gamma^a \psi^b\big) Y_{ac} P_\n{}^c{}_b  +\Delta\cL_3 \Big]  \ ,
%%%%%%%%%%%%%%%%%%%%%%%%%%%%%%%%%%%%%%%%%%%
\label{9}\w4
% old \circled{1} =
\circled{4} = &  ee^{2\hat\vp} \Big[ \tfrac12 \hat{R}_{\hat{\mu}\hat{\nu}}{}^{\hat{m}\hat{n}} ({\hat\Omega}_-)
\bar{\hat{\psi}}_{\hat{m}\hat{n}} \left( \hat{\gamma}^{\hat{\rho}}
\gamma^{\hat{\mu}\hat{\nu}} \hat{\psi}_{\hat{\rho}} 
+ 2 \hat{\gamma}^{\hat{\mu}\hat{\nu}} \hat{\chi} \right)\Big]  
\nn\w2
= &  ee^{2\vp} \Big[ \tfrac12 \big(\bpsi^\rh \gamma^{\m\n}\gamma_\rh -2{\bar\chi}\gamma^{\m\n} \big) \psi_{mn}\,R_{\m\n}{}^{mn} (\Omega_-)
+ \tfrac12\big( \bpsi^\rh\Gamma^{ab}\gamma_\rh  -2{\bar\chi}\Gamma^{ab}\big)  \psi_{mn}\, Q_{+}{}^{mn}{}_{ab} \nn\w2
&  +\tfrac12 \big(\bpsi^\rh \gamma^{\m\n}\gamma_\rh -2{\bar\chi}\gamma^{\m\n} \big)\gamma^\s \Gamma^c\psi_a\, \big(P_{\s cb} Q_{-\m\n}{}^{ab}\big)
\nn\w2
&  +\big(\bpsi^\rh\gamma_\rh \Gamma^{cd} -2{\bar\chi}\Gamma^{cd}\big) \Gamma^e \gamma^\nu \psi^a\, \big(P^\m{}_{ca} Y_{\m\n de}\big)
\nn\w2
&+ \big(\bpsi^\rh\gamma^\m\gamma_\rh \Gamma^b  + 2{\bar\chi}\gamma^\m \Gamma^b\big)\gamma^\n \Gamma^a \psi^\s\, \,\big(P_{\n ac} D_\m P_{\s b}{}{}^c  \big)
\nn\w2
&  +2 \big(\bpsi^\rh\gamma^\m\gamma_\rh \Gamma^b + 2{\bar\chi}\gamma^\m \Gamma^b\big)  \big(D^\n \psi^a\big) D_\m P_\n{}_{ba} +\Delta\cL_4 \Big]\ ,
%%%%%%%%%%%%%%%%%%%%%%%%%%%%%%%%%%%%%%%%%%%
\w4
% The terms inside square brackets were labeled originally as $\circled{3}, \circled{4}, \circled{5}+\circled{6}$ and $\circled{7} + \circled{8}$ 
\circled{5} =& ee^{2\hat\vp}\Big[ - \bar{\hat{\psi}}{}^{\hat{m}\hat{n}} \gamma^{\hat{\mu}} 
D_{\hat{\mu}}(\hat{\omega}, \hat{\Omega}_-) \hat{\psi}_{\hat{m}\hat{n}}
- \tfrac{1}{12} \hat{H}_{\hat{\mu}\hat{\nu}\hat{\rho}} 
\bar{\psi}^{\hat{m}\hat{n}} \hat{\gamma}^{\hat{\mu}\hat{\nu}\hat{\rho}} 
\hat{\psi}_{\hat{m}\hat{n}} \Big] 
\nn\w2
= & ee^{2\vp}\Bigg\{-\bar{\psi}^{mn} \gamma^\mu D_\mu(\omega, \Omega_-) \psi_{mn}
-\tfrac{1}{12} H_{\mu\nu\rho} \bar{\psi}^{mn} \gamma^{\mu\nu\rho} \psi_{mn} 
\nn\w2
& +\Big[ - \bar{\hat{\psi}}^{ab} \gamma^\mu D_\mu(\omega)\hat{\psi}_{ab}\Big] 
+\Big[ -2 \bar{\hat{\psi}}^{ma} \gamma^\mu D_\mu(\omega,\Omega_-)
\hat{\psi}_{ma}\Big]
\nn\w2
&  +\Big[-4\bar{\psi}^{mn} \Gamma^b \hat{\psi}_{na} P_{+m}{}^a{}_b 
-4 \bar{\hat{\psi}}^{ab} \Gamma^c \hat{\psi}_{mb} P_{+}{}^m{}_{ac}\Big]
+\Big[- \tfrac{1}{12} H_{\mu\nu\rho} \bigl(\bar{\hat{\psi}}^{ab} \gamma^{\mu\nu\rho} \hat{\psi}_{ab}
+ 2 \bar{\hat{\psi}}^{ma} \gamma^{\mu\nu\rho} \bpsi_{ma} \bigr) \Big] \Bigg\}\ ,
\nn\w4
=& ee^{2\vp} \Bigg\{ -\bar{\psi}^{mn} \gamma^\mu D_\mu(\omega, \Omega_-) \psi_{mn}
-\tfrac{1}{12} H_{\mu\nu\rho} \bar{\psi}^{mn} \gamma^{\mu\nu\rho} \psi_{mn} 
\nn\w2
& +\Big[ \tfrac12 \bpsi^b \g^\m \g^\rh \g^\n \Gamma^c \Gamma^d \big( D_\rh (\omega) \psi_b \big) ( P_{\m c}{}^a P_{\n d a} ) 
\nn\w2
& - \tfrac12 \bpsi^a \g^\m \g^\rh \g^\n \Gamma^c \Gamma^d ( D_\rh(\omega) \psi^b ) \big( P_{\m c b} P_{\n d a} \big)
\nn\w2
& + \tfrac12 \bpsi^b \g^\m \g^\rh \g^\n \Gamma^c \Gamma^d \psi_b ( P_{\m c}{}^a D_\rh(\Gamma) P_{\n d a}) 
- \tfrac12 \bpsi^a \g^\m \g^\rh \g^\n \Gamma^c \Gamma^d \psi^b \big(P_{\m c b} D_\rh(\Gamma) P_{\n d a} \big)\Big]
%%%%%%%%%%%%%%%%%%%%%%%%%%%%%%%%%%%%%%%%%
\nn\w4
& +\Big[ -2 ( D^m \bpsi^a ) \slashed{D} (\omega,\Omega_-) ( D_m \psi_a ) 
+ \bpsi^m \g^\m \Gamma^a ( \slashed{D} (\omega,\Omega_-)D_m \psi^b ) P_{\m a b} 
\nn\w2
&- ( D^m \bpsi^a ) \g^\m \g^\n \Gamma^b ( D_\m (\omega,\Omega_-) \psi_m ) P_{\n b a} 
- ( D^m \bpsi^a ) \g^\m \g^\n \Gamma^b \psi_m ( D_\m(\Gamma) P_{\n b a} ) 
\nn\w2
&+ \tfrac12 \bpsi^m \g^\n \g^\m \g^\rh \Gamma^b \Gamma^c ( D_\m (\omega,\Omega_-)  \psi_m ) P_{\rh c a} P_{\n b}{}^a  
+ \tfrac12 \bpsi^m \g^\m \g^\n \g^\rh \Gamma^b \Gamma^c \psi_m ( P_{\m b a} D_\n(\Gamma) P_{\rh c}{}^a ) 
\nn\w2
&  - 2 \bpsi^a \g^\m ( D^\n \psi^b ) Z_{\m\n ab} + \bpsi^\n \g^\m \g^\rh \Gamma^a \psi^b P_\rh{}^c{}_b Y_{\m\n ac} +\Delta\cL_5 \Big]
 %%%%%%%%%%%%%%%%%%%%%%%%    \Delta\cL_5 comes from $\circled{4}$    %%%%%%%%%%%%%%%%%%%%%%%
\nn\w4
& +\Big[ 4\bpsi^{mn} \Gamma_a \big( D_m \psi_b \big) P_n{}^{ab} 
- 2 \bpsi^{\n\rh} \g^\m \Gamma^a \Gamma^b \psi_\n Y_{\rh\m ab}
%%%%%%%%%%%%%%%%%%%%%%%%%%%%%%%%%%%%%%%%%
\nn\w4
&  - 2 \bpsi^a \g^\m \Gamma^b \Gamma^c ( D^\n \psi_a ) Y_{\m\n bc}  + 2 \bpsi^a \g^\m \Gamma^d \Gamma^c ( D^\n \psi^b ) P_{\m db} P_{\n ca} 
\nn\w2
&  - \bpsi^b \g^\m \g^\n \Gamma^d \Gamma^c \Gamma^e \psi^\rh \big( Y_{\m\rh dc} P_{\n eb} - Y_{\m\n de} P_{\rh cb}\big)  +\Delta\cL_6 \Big]
%%%%%%%%%%%%%%%    \Delta \cL_6 comes from \circled{5} + \circled{6}  %%%%%%%%%%%%%%%%%% 
\nn\w4
& +\Big[ \tfrac{1}{24} H_{\m\n\rh}\,\bpsi^b \Gamma^c\Gamma^d \gamma^\lambda \gamma^{\m\n\rh} \gamma^\tau \psi_b Y_{\lambda\tau cd} 
-\tfrac{1}{24} H_{\m\n\rh}\,\bpsi^a \Gamma^c\Gamma^d \gamma^\lambda \gamma^{\m\n\rh} \gamma^\tau \psi^b \big(P_{\lambda cb} P_{\tau da} \big)
%%%%%%%%%%%%%%%%%%%%%%%%%%%%%%%%%%%%%%%%
\nn\w4
& -\tfrac16 H_{\m\n\rh} \big(D^m \bpsi^a\big) \gamma^{\m\n\rh} \big(D_m\psi_a \big)
-\tfrac16 H_{\m\n\rh}\, \bpsi^m \Gamma^a \gamma^\s \gamma^{\m\n\rh} \big(D_m\psi^b\big) P_{\s ab}
\nn\w2
& +\tfrac{1}{24} H_{\m\n\rh}\, \bpsi^m \gamma^\la \gamma^{\m\n\rh}\gamma^\tau \Gamma^c \Gamma^d \psi_m Y_{\lambda\tau cd}  + \Delta\cL_7\Big] \Bigg\}\ .
%%%%%%%%%%%%%%%%%%%%%%%  \Delta \cL_7 comes from \circled{8}   %%%%%%%%%%%%%%%%%%%%%%%%%%%%% 
\label{Y1}
\end{align}
}
Furthermore, $D_\m\psi_a$ and $D_\m P_{\n ab}$ are as defined in \eq{dpsi} and \eq{DP2} and  
\begin{align}
D_\mu(\omega, \Omega_-) \psi_{rs} 
&= \left( \partial_{\mu} + \tfrac{1}{4} \omega_{\mu pq} \gamma^{pq} 
+ \tfrac{1}{4} Q_{+\mu ab} \Gamma^{ab}\right) \psi_{rs}
+ \Omega_{-\mu r}{}^p \psi_{ps} + \Omega_{-\mu s}{}^p \psi_{rp}\ ,
\nn\w2
D_\mu(\omega) \hat{\psi}_{ab} 
&= \left( \partial_{\mu} + \tfrac{1}{4} \omega_{\mu pq} \gamma^{pq} 
+ \tfrac{1}{4} Q_{+\mu cd} \Gamma^{cd} \right) {\hat\psi}_{ab}
+ Q_{-\mu a}{}^c {\hat\psi}_{cb} + Q_{-\mu b}{}^c {\hat\psi}_{ac}\ ,
\nn\w2
D_\mu(\omega, \Omega_-) \hat{\psi}_{ra} 
&= \left( \partial_{\mu} + \tfrac{1}{4} \omega_{\mu pq} \gamma^{pq} 
+ \tfrac{1}{4} Q_{+\mu cd}\Gamma^{cd}\right) {\hat\psi}_{ra}
+ \Omega_{-\mu r}{}^s {\hat\psi}_{sa} + Q_{-\mu a}{}^b {\hat\psi}_{rb}\ .
\end{align}
The terms in \eq{10plus11} can be absorbed into the CS terms in $H_{\mu\nu\rho}$ occurring in $\cL_0$, if we use supercovariant $\hat{\Omega}_-$ and $\hat{Q}_-$. The first term in \eq{9} can be absorbed into the $\mbox{Riem}(\Omega_-)^2$ term upon supercovariantization of $\Omega_-$. The two $10D$ terms in \eq{Y1} are reduced together for the following reason. The covariant derivative in the first term becomes $D_\mu(\omega, Q, \Omega_-, Q_-)$ in $6D$. When we add the $H_{\mu ab}$ contributions of the second term to the first term, the covariant derivative becomes $D_\mu(\omega,  Q_+, \Omega_-, Q_-)$, as $Q$ in the first term combines with $H_{\mu ab}$ in the second term and becomes $Q_+$. 
Summing all the $SO(4)_+ \times SO(4)_-$ breaking terms found above, while a large number of cancellations occur, we are still left with several terms given by 
\allowdisplaybreaks{
\begin{align}
\Delta \cL_F =& \Delta\cL_1+\Delta\cL_2+\Delta\cL_3+\Delta\cL_4+\Delta\cL_5 + \Delta\cL_6 + \Delta\cL_7
\nn\w2
=& 2\bpsi^\n \Gamma^a\psi^b \big(X_a{}^c P_{\n cb}\big)
+2\bpsi^\n\Gamma^a \psi^b \big( P^\m{}_a{}^c D_\m P_{\n cb}\big)
\nn\w2
& +2 \bpsi^\n\Gamma^{[a}\psi_c P_\n{}^{b]c} \,e^{-2\vp}D_\m(\Gamma)\big(e^{2\vp} P^\m{}_{ab}\big)
- 2  \bpsi^\n \Gamma^a\psi^b P_\n{}^c{}_b e^{-2\vp} D_\m \big( e^{2\vp} P^\m{}_{ac} \big) 
\nn\w2
& -  2 \bpsi^\n \Gamma^a\psi^b P^\m{}_a{}^c D_\m P_{\n cb}  
-2\big( \bpsi^\n\Gamma^a \psi^b\big) X_{ac} P_\n{}^c{}_b 
+2 \bpsi^a \g^\m \psi^b Y_a{}^c P_{\m cb}
%%%%%%%%%%%%%%%%%%%%%%%%%%%%%%%%%%%%%%%%%%%%%
\nn\w2
& +2\big( \bpsi_\rh \gamma^\m \gamma^\rh \Gamma^{(a} \psi^{b)} 
+ 2\bpsi^{(a}\Gamma^{b)} \gamma^\mu\chi\big)\big(P^\n{}_{ac} D_\m P_{\n b}{}^c \big)
\nn\w2
&  + \Big(  -\tfrac12 \bpsi_\n \gamma^{\m\n\rh} \Gamma^{ab} \psi_\rh 
-2\bpsi_\n \gamma^{\m\n} \Gamma^{ab} \chi -2{\bar\chi} \gamma^\m \Gamma^{ab} \chi 
+ \tfrac12 \bpsi^d \gamma^\m \Gamma^{ab} \psi_d  
\nn\w2
&   + 2\bpsi_\n \gamma^\m \gamma^\n \Gamma^{[a} \psi^{b]} 
-4\bpsi^{[a}\Gamma^{b]}  \gamma^\mu\chi + 2\bpsi_\n\gamma^\m\gamma^\n \Gamma^b \psi^a
+ 4\bpsi^a \Gamma^b \gamma^\m \chi \Big) Y_{ac} P_\m{}^{bc} 
%%%%%%%%%%%%%%%%%%%%%%%%%%%%%%%%%%%%%%%%%%
\nn\w2
& -2\bpsi^a\gamma^\m \big(D^\n \psi^b\big) D_\m P_{\n ab} 
+2 \big( \bpsi^\rh\gamma^\m\gamma_\rh \Gamma^b  + 2{\bar\chi}\gamma^\m \Gamma^b  \big)  \big(D_\n \psi^a\big) X_{\m\n ba}
\nn\w2
&  -2\bpsi^b\gamma^\m  \big(D_\n \psi^a\big) X_{\m\n ba}
 -  \bpsi^b\gamma^\m \gamma^\n \Gamma^a \psi^\rh\, \,\big(P_{\n ac} D_\m P_{\rh b}{}{}^c +P_{\rh b}{}^c D_\m P_{\n ac} \big)  
\nn\w2
& + \big(  \bpsi^\rh \gamma^\m \gamma_\rh \Gamma^b  +  2{\bar\chi}\gamma^\m \Gamma^b  \big)\gamma^\n \Gamma^d \psi^\s\, \big(P_{\mu bc}Y_{\s\n cd} \big)
 \nn\w2
& +\bpsi^b\gamma^\m \gamma^\n \Gamma^a \psi^\rh\, \big( -2P_{\m b}{}^c Y_{\n\rh ac}      +P_{\n a}{}^c X_{\m\rh bc} \big) 
%%%%%%%%%%%%%%%%%%%%%%%%%%%%%%%%%%%%%%%%%%%%%%%%%%%%%%
\nn\w2
& -2 \big(\bpsi^a \slashed{D}(\omega,\Omega_-) D^\m \psi^b \big) P_{\m a b}  - 2 ( D^\m \bpsi^a ) ( \slashed{D}(\omega) \psi^b)  P_{\m b a} 
\nn\w2
&  - \bpsi^b \g^\m \g^\n \Gamma^c \big( D_\m (\omega,\Gamma_+)\psi^\rh \big)  Y_{\n\rh cb} 
- 2 \bpsi^a( \slashed{D}(\omega) \psi^b ) Y_{ab} 
\nn\w2
& - \bpsi^\n \Gamma^a \g^\m (\slashed{D}(\omega) \psi^b ) Y_{\m\n ab}
  - \bpsi^\n \g^{\m\rh} \Gamma^a \psi^b H_{\m\rh}{}^\ta Y_{\ta\n a b}
%%%%%%%%%%%%%%%%%%%%%%%%%%%
\nn\w2
& -\tfrac16 H_{\m\n\rh} \Big( 2\bpsi^a \gamma^{\m\n\rh} \big(D_\s \psi^b \big) P^\s{}_{ab}
+\bpsi^b \gamma^{\m\n\rh} \gamma^\s \Gamma^a \psi^\tau Y_{\s\tau ab}
+\bpsi^a \gamma^{\m\n\rh} \psi^b Y_{ab} \Big)\ .
\label{DLF}
\end{align}}

%%%%%%%%%%%%%%%%%%%%%%%%%%%%%%%%%%%%%%%%%%%%%%%%%%%%%%%%%%%%%%%%%%%%%%%%%%%%%
\section{Field redefinitions and the total Lagrangian}
%%%%%%%%%%%%%%%%%%%%%%%%%%%%%%%%%%%%%%%%%%%%%%%%%%%%%%%%%%%%%%%%%%%%%%%%%%%%%

Combining the  results  for ${\Delta \cL}_B$ and ${\Delta\cL}_F$ given \eq{deltaLB} and \eq{DLF}, respectively, with the Lagrangian generated by the field redefinitions in $\cL_0$ described in  Appendix B,  most remarkably we find that 
\begin{align}
& ee^{2\vp} \big( \Delta\cL_B + \Delta\cL_F \big) + \Delta \cL_0 \Big|_{E \to YE}   +\delta \cL_0 \Big|_{\psi^a \to \psi^\m Y} +\delta \cL_0 \Big|_{\psi \to Y D \psi} 
+\delta\cL_0 \Big|_{\psi \to Y\psi} +\delta \cL_0\Big|_{E\to \psi_\n \psi_a PE} 
\nn\w2
& =  ee^{2\vp} \Big[ -Z_{\m\n ab}Z^{\m\n ab}
+\Big(\bpsi_\n \g^\m \g^\n \Gamma^a \psi^b + 2 \bar{\chi} 
\g^\m \Gamma^a \psi^b \Big) P_\m{}^c{}_b Y_{ac} -\bpsi^a \gamma^\rh\gamma^\m \Gamma^b \psi^\n 
\big( P_\rh{}^c{}_a Y_{\m\n bc} \big) 
\nn\w2
&  -2\bpsi^a\gamma^\m \big(D^\n \psi^b \big) Z_{\m\n ab} \Big] \ .
\label{ffd}
\end{align}
All terms displayed in \eq{DLF}, which are only invariant under the diagonal $SO(4)$ subgroup, have cancelled as a result of the field redefinitions, and a handful new terms are produced that are invariant under $SO(4)_+\times SO(4)_-$.
%
%%%%%%%%%%%%%%%%      THE TOTAL DESIRED LAGRANGIAN     %%%%%%%%%%%%%%%%%%%%%%%
%
Putting together all the results described above we obtain 
\allowdisplaybreaks{
\begin{align}
\cL\Big|_\OA = & \cL_B \Big|_\OA + \circled{1} +\cdots +\circled{5} +\del \cL_0 \Big|_{E \to YE}   +\delta \cL_0 \Big|_{\psi^a \to \psi^\m Y} +\delta \cL_0 \Big|_{\psi \to Y D \psi} 
\nn\w2
&  +\delta \cL_0 \Big|_{\psi \to Y\psi} +\delta \cL_0\Big|_{E\to \psi_\n \psi_a PE} 
%%%%%%%%%%%%%%%%%%%%%%%%%%%%%%%%%%
\nn\w2
= & \alpha' e e^{2\varphi}\Bigg\{ 
\Big[  H^{\m\n\rh} \big( \omega^L_{\m \n \rh}(\Omega_-) +\omega^Q_{\m\n\rh} (Q_-) \big) 
- \tfrac14 R_{\mu\nu rs}(\Omega_-) R^{\mu\nu rs}(\Omega_-) - \tfrac14  Q_{+\mu\nu ab} Q_+{}^{\mu\nu ab}
\nn\w2
& - \tfrac14  Q_{-\mu\nu ab} Q_-{}^{\mu\nu ab}
-D_\mu P_{\n ab}  D^\mu  P^{\n ab} 
-\tfrac12 Y_{\m\n} Y^{\m\n} +\tfrac12 Z_{\m\n ab} Z^{\m\n ba} -Z_{\m\n ab}Z^{\m\n ab} \Big]
\nn\w2
& + \Big[  -H^{\m\n\rh} R_{\m\n}{}^{rs} (\Omega_-)\,\bpsi_r \gamma_\rh \psi_s - H^{\m\n\rh} Q_{-\m\n}{}^{ab}\,\bpsi_a \gamma_\rh \psi_b
\nn\w2
&  +\big( \bpsi_r\gamma_\n\psi_s \Omega_{-\rh}{}^{rs} + \bpsi_a\gamma_\n\psi_b Q_{-\rh}{}^{ab}\big) e^{-2\vp} D_\m(\Gamma)\big(e^{2\vp} H^{\m\n\rh}\big) \Big]
%%%%%%%%%%%%%%%%%%%%%%%%%%%%%%%%%%%%%%%%%%%%%%%%%%%%
\nn\w4
& +\Big[ \tfrac14 \big( \omega^L_{\m\n\rh}(\Omega_-) + \omega^Q_{\m\n\rh}(Q_-) \big)   \left( 
\bar{\psi}^\sigma \gamma_{[\sigma} \gamma^{\mu\nu\rho} 
\gamma_{\tau]} \psi^\tau  + 4 \bar{\psi}_\sigma \gamma^{\sigma\mu\nu\rho} \chi 
- 4 \bar{\chi} \gamma^{\mu\nu\rho} \chi   +\bpsi^a \gamma^{\m\n\rh} \psi_a  \right)
\nn\w2
& +\tfrac12 \Big( -\bpsi_\n \gamma^{\m\n\rh} \Gamma^{ab} \psi_\rh 
-4\bpsi_\n \gamma^{\m\n} \Gamma^{ab} \chi -4{\bar\chi} \gamma^\m \Gamma^{ab} \chi + \bpsi^d\gamma^\m \Gamma^{ab}  \psi_d \Big) P^\s{}_a{}^c D_\m P_{\s bc} \Big]
%%%%%%%%%%%%%%%%%%%%%%%%%%%%%%%%%%%%%%%%%
\nn\w4
& +\Big[ -2 R^{\m\n rs}(\Omega_-) \bpsi_r \gamma_\n D_\m (\omega,\Omega_-)\psi_s
-2 Q_{-\m\n}{}^{ab}  \bpsi_a \gamma^\n D^\m (\omega) \psi_b
+2\bpsi_\m\Gamma_b\psi^c \Big(Q_+{}^{\m\n ab} P_{\n ac}\Big)
\nn\w2
& +2 \Big(\bpsi^r\Gamma^a D_\m (\omega)\psi^b -\bpsi^b\Gamma^a D_\m(\omega,\Omega_-)\psi^r\Big) D^\m P_{r ab}
-2\bpsi_r\gamma_\m\psi_\n \Big(P^\n{}_{ab} D_\m P_r{}^{ab}\Big)
\nn\w2
&  -2\bpsi_b\gamma_\m\psi_c \Big(P^{\n ac}  D_\m P_{\n a}{}^b\Big)   + 2 \bpsi^\n \Gamma^a \psi^b  P^\m{}_{ab} Y_{\m\n} -2\bpsi^\n \Gamma^a \psi^b \big(P^{\m c}{}_b Y_{\m\n ac}\big) + 2 \bpsi^\n \Gamma^a \psi^b (Y_{ac} P_\n{}^c{}_b) \Big]
%%%%%%%%%%%%%%%%%%%%%%%%%%%%%%%%%%%%%%%%%%%
\nn\w4
& +\Big[\tfrac12 \big(\bpsi^\rh \gamma^{\m\n}\gamma_\rh -2{\bar\chi}\gamma^{\m\n} \big) \psi_{mn}\,R_{\m\n}{}^{mn} (\Omega_-)
+ \tfrac12\big( \bpsi^\rh\Gamma^{ab}\gamma_\rh  -2{\bar\chi}\Gamma^{ab}\big)  \psi_{mn}\, Q_{+}{}^{mn}{}_{ab} \nn\w2
& +\tfrac12 \big(\bpsi^\rh \gamma^{\m\n}\gamma_\rh -2{\bar\chi}\gamma^{\m\n} \big)\gamma^\s \Gamma^c\psi_a\, 
\big(P_{\s cb} Q_{-\m\n}{}^{ab}\big)
+\big(\bpsi^\rh\gamma_\rh \Gamma^{cd} -2{\bar\chi}\Gamma^{cd}\big) \Gamma^e \gamma^\nu \psi^a\, 
\big(P^\m{}_{ca} Y_{\m\n de}\big)
\nn\w2
& + \big(\bpsi^\rh\gamma^\m\gamma_\rh \Gamma^b  + 2{\bar\chi}\gamma^\m \Gamma^b\big)\gamma^\n \Gamma^a \psi^\s\, \big(P_{\n ac} D_\m P_{\s b}{}{}^c  \big)
+2 \big(\bpsi^\rh\gamma^\m\gamma_\rh \Gamma^b + 2{\bar\chi}\gamma^\m \Gamma^b\big)  \big(D^\n \psi^a\big) D_\m P_\n{}_{ba} \Big]
%%%%%%%%%%%%%%%%%%%%%%%%%%
\nn\w4
& - \bar{\psi}^{mn} \gamma^\mu D_\mu (\omega,\Omega_-)\,  \psi_{mn}
- \frac{1}{12} H_{\mu\nu\rho} 
\bar{\psi}^{mn} \gamma^{\mu\nu\rho} \psi_{mn}
\nn\w2
& +\Big[ \tfrac12 \bpsi^b \g^\m \g^\rh \g^\n \Gamma^c \Gamma^d \big( D_\rh (\omega) \psi_b \big) ( P_{\m c}{}^a P_{\n d a} ) - \tfrac12 \bpsi^a \g^\m \g^\rh \g^\n \Gamma^c \Gamma^d ( D_\rh(\omega) \psi^b ) \big( P_{\m c b} P_{\n d a} \big)
\nn\w2
& + \tfrac12 \bpsi^b \g^\m \Gamma^{cd}\psi_b ( P_{\m c}{}^a D^\n P_{\n d a}) - \tfrac12 \bpsi^a \g^\m \Gamma^c \Gamma^d \psi^b \big(P_{\m c b} D^\n  P_{\n d a} \big) \Big]
%%%%%%%%%%%%%%%%%%%%%%%%%%%%%%%%%%%%%%%%%
\nn \w4
& +\Big[ - 2 ( D^m \bpsi^a ) \slashed{D} (\omega,\Omega_-) ( D_m \psi_a ) + P_{\m a b}\,\bpsi^m \g^\m \Gamma^a  \slashed{D} (\omega,\Omega_-)D_m \psi^b   
\nn\w2
&- P_{\n b a} \,( D^m \bpsi^a ) \g^\m \g^\n \Gamma^b  D_\m (\omega,\Omega_-) \psi_m  
- ( D^m \bpsi^a ) \Gamma^b \psi_m  D^\n P_{\n b a} 
\nn\w2
&+ \tfrac12 Y_{\n\rh ab}\, \bpsi^m \g^\n \g^\m \g^\rh \Gamma^a \Gamma^b  D_\m (\omega,\Omega_-)  \psi_m   + \tfrac12 \bpsi^m \g^\m  \Gamma^{bc} \psi_m ( P_{\m b a} D^\n  P_{\n c}{}^a )
\nn\w2
&  - 2Z_{\m\n ab}\, \bpsi^a \g^\m D^\n \psi^b  + \bpsi^\n \g^\m \g^\rh \Gamma^a \psi^b P_\rh{}^c{}_b Y_{\m\n ac} \Big]
%%%%%%%%%%%%%%%%%%%%%%%%%%%%%%%%%%%%%%%%
\nn\w4
& +\Big[ 4P_n{}^{ab}\, \bpsi^{mn} \Gamma_a D_m \psi_b 
- 2 \bpsi^{\n\rh} \g^\m \Gamma^a \Gamma^b \psi_\n Y_{\rh\m ab}
- 2Y_{\m\n bc}\, \bpsi^a \g^\m \Gamma^b \Gamma^c  D^\n \psi_a
%%%%%%%%%%%%%%%%%%%%%%%%%%%%%%%%%%%%%%%%%
\nn\w4
&  + 2 \bpsi^a \g^\m \Gamma^d \Gamma^c ( D^\n \psi^b ) P_{\m db} P_{\n ca}  - \bpsi^b \g^\m \g^\n \Gamma^d \Gamma^c \Gamma^e \psi^\rh \big( Y_{\m\rh dc} P_{\n eb} - Y_{\m\n de} P_{\rh cb}\big) 
\Big]
%%%%%%%%%%%%%%%%%%%%%%%%%%%%%%%%%%%%%%%%%
\nn\w4
& +\Big[ \tfrac{1}{24} H_{\m\n\rh}\,\bpsi^b \Gamma^c\Gamma^d \gamma^\lambda \gamma^{\m\n\rh} \gamma^\tau \psi_b Y_{\lambda\tau cd} 
-\tfrac{1}{24} H_{\m\n\rh}\,\bpsi^a \Gamma^c\Gamma^d \gamma^\lambda \gamma^{\m\n\rh} \gamma^\tau \psi^b \big(P_{\lambda cb} P_{\tau da} \big)
%%%%%%%%%%%%%%%%%%%%%%%%%%%%%%%%%%%%%%%%
\nn\w4
& -\tfrac16 H_{\m\n\rh} \big(D^m \bpsi^a\big) \gamma^{\m\n\rh} \big(D_m\psi_a \big)
-\tfrac16 H_{\m\n\rh} P_{\s ab}\, \bpsi^m \Gamma^a \gamma^\s \gamma^{\m\n\rh} D_m\psi^b
\nn\w2
& +\tfrac{1}{24} H_{\m\n\rh} Y_{\lambda\tau cd}\, \bpsi^m \gamma^\la \gamma^{\m\n\rh}\gamma^\tau \Gamma^c \Gamma^d \psi_m \Big]
%%%%%%%%%%%%%%%%%%%%%%%%%%%%%%%%%%%%%%%%
\nn\w4
& +\Big[ \Big(\bpsi_\n \g^\m \g^\n \Gamma^a \psi^b + 2 \bar{\chi} 
\g^\m \Gamma^a \psi^b \Big) P_\m{}^c{}_b Y_{ac} -\bpsi^a \gamma^\rh\gamma^\m \Gamma^b \psi^\n 
\big( P_\rh{}^c{}_a Y_{\m\n bc} \big)
\nn\w2
& -2Z_{\m\n ab}\,\bpsi^a\gamma^\m D^\n \psi^b  \Big]
\Bigg\} \ .
\label{ML1}
\end{align}}
The term $-ee^{2\vp} Z_{\m\n ab} Z^{\m\n ab}$ and the last four terms arise from the field redefinitions. The sum of $\cL_0$ and the $\OA$ Lagrangian above can be simplified by the modification of the 3-form field strength by Chern-Simons terms and various supercovariantizations. This is done in Appendix C, where the terms in the total Lagrangian are grouped in a systematic way according to their structures.

\subsection{Supersymmetry transformations at $\OA$ }
%%%%%%%%%%%%%%%%%%%%%%%%%%%%%%%%%%%%%%%%%%%%%%%%%%%%%%%%%%%%%%%%%%%%

The dimensional reduction of the supersymmetry transformations at lowest order in $\alpha'$ is given in \eq{susy1}. Here, we shall determine the supersymmetry transformations at $\OA$. In doing so we shall also take into account the field redefinitions discussed in Appendix B.

Prior to the field redefinitions, the dimensional reduction up to ${\cal O}(\alpha')$ gives the supertransformations to cubic terms in fermions as 
\begin{align}
\delta e_\mu{}^r &= \bar{\epsilon} \gamma^r \psi_\mu\ , 
\nn\w2
\delta \psi_\mu &= D_\mu(\Omega_+) \epsilon  -\tfrac32 \alpha' \big[ \omega^L_{\m\n\rh}(\Omega_-) + \omega^Q_{\m\n\rh}(Q_-) \big] \gamma^{\n\rh} \e 
\nn\w2
& - \alpha' P_{\n a}{}^c \big( D_\m P^\n{}_{bc} + X_\m{}^\n{}_{bc} \big)\Gamma^{ab} \e\ , \
\nn\w2
\delta B_{\mu\nu} 
&= - \bar{\epsilon} \gamma_{[\mu} \psi_{\nu]} +2\alpha' \Omega_{-[\m}{}^{rs} \delta_0 \Omega^{(sc)}_{-\n] rs} +2\alpha' Q_{-[\m}{}^{ab} \delta_0 Q^{(sc)}_{-\n]ab}  \ , 
\nn\w2
\delta\chi 
&= \tfrac{1}{2} \gamma^\mu \epsilon \partial_\mu  \varphi 
- \tfrac{1}{12} H_{\mu\nu\rho} \gamma^{\mu\nu\rho} \epsilon  +\tfrac12 \alpha' \big[ \omega^L_{\m\n\rh}(\Omega_-) + \omega^Q_{\m\n\rh}(Q_-) \big] \gamma^{\m\n\rh} \e \ , 
\nn\w2
\delta\varphi &= \bar{\epsilon} \chi\ , 
\nn\w2
 W \delta W^{-1} &= 
\left(
\begin{array}{c|c}
 -2{\bar\e}\Gamma_{[a} \psi_{b]} + 4\alpha' P_{[a}{}^c \delta P^\m{}_{b]c} \  & 
 \ -{\bar\e}\Gamma_a \psi_b + 4\alpha' P_{[a}{}^c \delta P^\m{}_{b]c} \\
\hline
\  -{\bar\e}\Gamma_b \psi_a + 4\alpha' P_{[b}{}^c \delta P^\m{}_{a]c}  & -4\alpha' P_{[a}{}^c \delta P^\m{}_{b]c}  \ 
\end{array}
\right)\ ,
%\label{dw}
\nn\w2
\delta \psi_a 
&= - \tfrac12 \g^\m \Gamma^b \e P_{\m b a} +2\alpha'P_{\n a}{}^c \big( D_\m P^\n{}_{bc} + X_\m{}^\n{}_{bc} \big) \Big|_{[ab]}\gamma^\m \Gamma^b \epsilon \ ,
\label{susy2}
\end{align}
where
\be
\Omega^{(sc)}_{-\mu rs} = \Omega_{-\mu rs} + \bar{\psi}_r \gamma_\mu \psi_s\ , 
\qquad
Q^{(sc)}_{-\mu ab} = Q_{-\mu ab} + \bar{\psi}_a \gamma_\mu \psi_b\ .
\label{scc}
\ee
For later purposes, let us also record the transformation of $B_{\m\n}$ under the $SO(4)_{-}$ transformations:
\be
\delta_\Lambda B_{\m\n} = 2\alpha'\Lambda_{-}^{ab} \partial_{[\m} Q_{-\n] ab}\ .
\label{BT}
\ee
Performing the field redefinition $E_\alpha{}^a = E'_\alpha{}^a + \delta E_\alpha{}^a$ with $\delta E_\alpha{}^a = -2\alpha' Y^a{}_b E_\alpha{}^b$, and noting the formula \eq{dpq} that gives
\be
Q_{+\m ab} = Q'_{+\m ab} -4\alpha' P_{\m [a}{}^c Y_{b]c}\ ,
\ee
we find that
\be
\delta \psi_\mu = D_\mu(\Omega_+, Q'_+) \epsilon  -\tfrac32 \alpha' \big[ \omega^L_{\m\n\rh}(\Omega_-) + \omega^Q_{\m\n\rh}(Q_-) \big] \gamma^{\n\rh} \e   - \alpha' P_{\n a}{}^c D_\m P^\n{}_{bc} \Gamma^{ab} \e\ .
\ee
In the  last two terms $Q_-$ and $P$ can be primed since we are interested only at $\OA$ terms. Next, considering the redefinitions  $\psi_a$ specified in Appendix B as well, we have the redefined fields
\begin{align}
\psi_a' =&\, \psi_a + 2 \al' \psi^b Y_{ab} + 2 \al' \g^\n \Gamma^b \psi^\m Y_{\m\n ab} +4 \al' P_{\m a b} D^\m(\Omega_+) \psi^b \ , \label{fr1}
\w2
E'_\al{}^a =&\, E_\al{}^a + 2 \al' E_{\al b} Y^{ab} -4\alpha'\, \bpsi^\n \Gamma^{(a} \psi_c P_\n{}^{b)c} E_{\alpha b}\ . 
\label{fr2}
\end{align}
It is noteworthy that the third term in \eq{fr1} supercovariantizes the derivative in the last term, and the third term in \eq{fr2} supercovariantizes $Y_{ab}$ up to quartic fermion terms. It follows from \eq{fr1} that 
\begin{equation}
\begin{aligned}
\del \psi'_a =&\,  - \tfrac12 \g^\m \Gamma^b \e P_{\m b a} - \al' \g^\m \Gamma^b \e P_{\m b}{}^c Y_{ac} - 2 \al' \g^\m \Gamma^b \e P_{\n a}{}^c D_\m P^\n{}_{bc} 
\w2
&\,+ 2 \al' P_{\n a}{}^c \big( D_\m P^\n{}_{bc} + X_\m{}^\n{}_{bc} \big) \Big|_{[ab]}\gamma^\m \Gamma^b \epsilon \ .
\end{aligned}
\label{nt1}
\end{equation}
In the last three terms $P_{\m ab}$, and therefore $Y_{ab}$ by  $X_{\m\n bc}$, can be primed since we are interested in $\OA$ terms. Next, using 
\begin{equation}
\begin{aligned}
P'_{\m a b} =&\, P_{\m a b} + 2 \al' D_\m Y_{ab} + 2 \al' ( P_{\m b}{}^c - P_\m{}^c{}_b ) Y_{ac}\ ,
\end{aligned}
\label{fr3}
\end{equation}
in the first term of \eq{nt1}, a number of $SO(4)_+ \times SO(4)_-$ symmetry breaking terms cancel out, and we end up with the $\OA$ result invariant under $SO(4)_+ \times SO(4)_-$ given by
\begin{align}
\del \psi'_a =&  - \tfrac12 \g^\m \Gamma^b \e  P^\prime_{\m ba} - \al'\g^\m \Gamma^b \e  P^\prime_\m{}^c{}_a  Y^\prime_{bc}\ . 
\label{nt2}
\end{align}

Turning to the supertransformation of the hyperscalars, 
\be
W' \delta W^{'-1} = 
\left(
\begin{array}{c|c}
 E'_{[a|}{}^\alpha \delta E'_{\alpha |b]} +E'_a{}^\alpha E'_b{}^\beta \delta B_{\alpha\beta} \  & 
 \ -E'_{(a|}{}^\alpha \delta E'_{\alpha |b)} +E'_a{}^\alpha E'_b{}^\beta \delta B_{\alpha\beta} \\
\hline
\ -E'_{(a|}{}^\alpha \delta E'_{\alpha |b)} +E'_b{}^\alpha E'_a{}^\beta \delta B_{\alpha\beta}  & E'_{[a|}{}^\alpha \delta E'_{\alpha |b]} -E'_a{}^\alpha E'_b{}^\beta \delta B_{\alpha\beta} \ 
\end{array}
\right)\ ,
\label{dw}
\ee
where we recall that $B_{\alpha\beta}$ does not undergo any field redefinition.  The supertransformation of \eq{fr2} up to $\OA$ yields 
\bea
\delta E'_{\alpha b} &=& E_\al{}^a \Big(\, {\bar\e}\Gamma_b \psi_a - 2\alpha' Y_{ac}\,{\bar\e} \Gamma^c \psi_b + 4 \al' \bar{\e} \Gamma_{(a} \psi^c Y_{b)c} 
\nn\w2
&&\, + 4 \al' P_{\m (a}{}^c \bar{\e} \Gamma_{b)} D^\m \psi_c - 2 \al' \bar{\e} \g^\m \Gamma_{(a|} \Gamma^c \psi^\n Y_{\m\n c |b)}  \Big)\ ,
\eea
while the reduction of the $10D$ supertransformations gives
\begin{equation}
\delta B_{\alpha\beta} = E_{[\alpha}{}^a E_{\beta]}{}^b \Big( - {\bar\e}\Gamma_a \psi_b -4\alpha' P^\m{}_b{}^c {\bar\e} \Gamma_a D_\m \psi_c -4\alpha' Y_b{}^c {\bar\e}\Gamma_a\psi_c  +2\alpha'Y_{\m\n cb} {\bar\e} \gamma^\m \Gamma_a \Gamma^c \psi^\n \Big)\ .
\end{equation}
Passing over to the primed fields, and up to $\OA$, the last two supertransformations take the form 
\bea
\del E'_\al{}^b &=& E'_{\al a} \Big[ \bar{\e} \Gamma^b \psi'^a - 2 \al' \bar{\e} \Gamma_c \psi'^b Y'^{ac} + 4 \al' \bar{\e} \g^\m \psi^\n Y'_{\m\n}{}^{[ba]}  + 4 \al' \bar{\e} \Gamma^{[a|} ( D_\m \psi'_c ) P'^{\m |b] c} 
\nn\w2
&& + 4 \al' \bar{\e} \Gamma^{[a} \psi'_c Y'^{b] c} + 2 \al' \bar{\e} \g^\n \Gamma^c \Gamma^{[a|} \psi^\m Y'_{\m\n}{}^{|b]}{}_c \Big]\ , 
\w2
\del B_{\al\beta} &=& E'_{[\al}{}^a E'_{\beta]}{}^b \Big[ - \bar{\e} \Gamma_a \psi'_b + 2 \al' \bar{\e} \Gamma_c \psi'_b Y_a{}^c \Big] \ .
\eea
In \eq{dw} the terms in the upper and lower block on the diagonal can be removed by $SO(4)_+$ and $SO(4)_-$ gauge transformations respectively. As for the remaining components in \eq{dw}, using the results for $\delta E'_\alpha$ and $\delta B_{\al\beta}$ in \eq{dw}, we obtain up to $\OA$ the supertransformation
\be
W' \big( \delta + \delta_{SO(4)_+} + \delta_{SO(4)_-}\big) W'^{-1} = \left(
\begin{array}{c|c}
 0 \  & 
 \ - \bar{\e} \Gamma_a \psi'_b + 2 \al' \bar{\e} \Gamma_c \psi'_b Y'_a{}^c 
\\
\hline
\ - \bar{\e} \Gamma_b \psi'_a + 2 \al' \bar{\e} \Gamma_c \psi'_a Y'_b{}^c   & 
0 \ 
\end{array}
\right)\ .
\ee
The lower right block of \eqref{dw} is
\begin{align}
\Lambda_{-ab} =& E'_{[a|}{}^\alpha \delta E'_{\alpha |b]} -E'_a{}^\alpha E'_b{}^\beta \delta B_{\alpha\beta}
\nn\\
=&\,- 4 \al' \bar{\e} \Gamma^c \psi'_{[b} Y'_{a]c} + 4 \al' \bar{\e} \Gamma_{[a} \psi'^c Y'_{b]c} 
+ 4 \al' \bar{\e} \Gamma_{[a|} ( D_\m \psi'_c ) P'^\m{}_{|b]}{}^c  
\nn\w2
&\, - 2 \al' \bar{\e} \g^\n \Gamma_{[a|} \Gamma^c \psi^\m Y'_{\m\n|b]c} \ .
\end{align}
The upper left block of \eqref{dw} is
\begin{align}
\Lambda_{+ab} =& E'_{[a|}{}^\alpha \delta E'_{\alpha |b]} + E'_a{}^\alpha E'_b{}^\beta \delta B_{\alpha\beta} \nn\\
=&\, 2 \bar{\e} \Gamma_{[b} \psi'_{a]} + 4 \al' \bar{\e} \Gamma_{[a} \psi'^c Y'_{b]c} 
 + 4 \al' \bar{\e} \Gamma_{[a|} ( D_\m \psi'_c ) P'^\m{}_{|b]}{}^c
 \nn\w2
&\, - 2 \al' \bar{\e} \g^\n \Gamma_{[a|} \Gamma^c \psi^\m Y'_{\m\n|b]c} \ .
\end{align}
The compensating $SO(4)_{+}$ transformation acts on fermions thereby giving higher order in fermion terms which we are neglecting. As for the compensating $SO(4)_{-}$ transformations, they act on $B_{\m\n}$ but giving rise to to quadratic in $\alpha'$ terms, which we are also neglecting.

\subsection{Closer look at the bosonic action}
%%%%%%%%%%%%%%%%%%%%%%%%%%%%%%%%%%%%%%%%%%%%%%%%%%%%%%%%%%%%%%%%%%%

Let us have a closer look at the bosonic part of this Lagrangian, which we denote by $\cL_{Bos., \OA}$. Noting that
\begin{align}
Q_{+\m\n ab} =& -2(P_\m P_\n^T )_{[ab]}\ ,\ \ Q_{-\m\n ab} = -2(P^T_\m P_\n)_{[ab]}\ ,\ \ Z_{\m\n ab} = (P_\m^T P_\n)_{ab}\ ,
\nn\w2
 Y_{\m\n} =&  \tr(P_\m P_\n^T)\ ,
\end{align}
it can be written as
\begin{align} 
\cL_{Bos., \OA} =& e e^{2\varphi}
\Big[  H^{\m\n\rh} \big( \omega^L_{\m\n\rh}(\Omega_-)+\omega^Q_{\m\n\rh} (Q_-)\big)
- \tfrac14 R_{\mu\nu mn}(\Omega_-) R^{\mu\nu mn}(\Omega_-) 
\nn\w2
&  -\tr ( D_\m (\Gamma_+)P_\n D^\m(\Gamma_+) P^{\n T} ) - \tfrac12 \tr ( P_\m P_\n^T ) \tr ( P^\m P^{\n T} ) 
\nn\w2
&\, - \tfrac32 \tr ( P_\m P^{\m T} P_\n P^{\n T} ) - \tfrac12 \tr ( P_\m^T P^\m P_\n^T P^\n ) 
\nn\w2
&\, + \tfrac32 \tr ( P_\m P_\n^T P^\m P^{\n T} ) 
\Big]\ .
\label{LB2}
\end{align}
To compare this result with that of \cite{Eloy:2020dko}, we need to evaluate it on the $\cL_0$-shell. To begin with, using the fact that
\be
R_{\m\n\rh\s}(\Gamma_+) = R_{\m\n\rh\s} (\Gamma) - 2D_{[\m}(\Gamma) H_{\n]\rh\s} - 2 H_{\m\rh,\n\s}\ ,
\ee
where $H_{\m\n,\rh\s} := H_{\m\n\tau} H_{\rh\s}{}^\tau$, we find the following relations 
\begin{align}
& \int ee^{2\vp} R_{\m\n\rh\s}(\Gamma_+) R^{\m\n\rh\s}(\Gamma_+) = \int ee^{2\vp} \Big[ R_{\m\n\rh\s}(\Gamma) R^{\m\n\rh\s}(\Gamma) +2 R_{\m\n\rh\s}(\Gamma) H^{\m\n,\rh\s} 
\nn\w2
&\qquad\qquad  - 2 H_{\m\n,\rh\s} H^{\m\rh,\n\s} -2 H^2_{\m\n} H^{2 \m\n} -4 H^{2 \m\n} \tr (P_\m P_\n^T) 
-16 H^2_{\m\n} \mathcal{E}^{\m\n} 
\nn\w2
&\qquad \qquad + \big( -4H^2 \mathcal{E}_\vp +8H_{\m\n\rh} (\partial^\m\vp) \mathcal{B}^{\n\rh} -4 H_{\m\n\rh} D^\m \mathcal{B}^{\n\rh}) \Big]\ ,
\label{lemma1}
\w4
& - \int e\, e^{2 \vp} \tr \Big( \left( D_\m( \Gamma_+) P_\n \right) \left( D^\m(\Gamma_+) P^{T\n} \right) \Big)  =\, \int e\, e^{2 \vp} \Big[  \tr (P_\m P^T_\n) \tr (P^\m P^{T\n})  
\nn\w2
& -\tr \big(P_\m P^T_\n P^\m P^{T\n}\big) + \tr \big( P_\m^T P^\m P^T_\n P^\n \big) 
-\tr \big( P^T_\m P_\n P^{T\m} P^\n \big) + \tr \big(P^\m P_\m^T P^\n P_\n^T \big)
\nn\w2
& \qquad\qquad +  \big(  4 \mathcal{E}_{\m\n} + \mathcal{E}_\vp g_{\m\n} \big) \tr ( P_\m P^T_\n)
+ ( D_\n \mathcal{E}_P^{ab} ) P^\n{}_{ab} - 2  \mathcal{E}_P^{ab} P^\n{}_{ab}  \pd_\n \vp \Big]\ ,
\label{lemma2}
\w4
&  \int H^{\m\n\rh} \omega^L_{\m \n \rh} (\Omega_-)= \int \big[ H^{\m\n\rh}  \omega^L_{\m \n \rh} (\omega) + R_{\m\n\rh\s} (\Gamma) H_{\m\n,\rh\s} -\tfrac23 H_{\m\n,\rh\s} H_{\m\rh,\n\s} \big]\ ,
\label{lemma3}
\end{align}
where $H^2_{\m\n} := H_{\m\rh\s} H_\n{}^{\rh\s}$ and the field equations that follow from $\cL_0$ are given in \eqref{bosfe}. Using these relations in \eq{LB2} we find the on $\cL_0$-shell result
\begin{align} 
\cL_{Bos., \OA} =& e e^{2\varphi}
\Big[ H^{\m\n\rh} \big( \omega^L_{\m \n \rh} (\omega) +\omega^Q_{\m\n\rh}(Q_-) \big)
 - \tfrac14 R_{\mu\nu mn}(\omega) R^{\mu\nu mn}(\omega) 
\nn\w2
& +\tfrac12 R_{\m\n\rh\s} H^{\m\n,\rh\s}  +\tfrac12 H^2_{\m\n} H^{2 \m\n} - \tfrac16 H_{\m\n,\rh\s} H^{\m\rh,\n\s} +  H^{2\m\n} \tr (P_\m P_\n^T)  
\nn\w2
& +  \tfrac12 \tr (P_\m P^T_\n) \tr (P^\m P^{T\n})   -\tfrac12\tr \big(P_\m P^T_\n P^\m P^{T\n}\big) +\tfrac12 \tr \big( P_\m^T P^\m P^T_\n P^\n \big) 
\nn\w2
&  -\tfrac12 \tr \big(P^\m P_\m^T P^\n P_\n^T \big) \Big]\ .
\label{LB3}
\end{align}
Comparison of this result with that of \cite{Eloy:2020dko} requires the introduction of the $O(4,4)$ matrix
\be
{\cal S} = \eta V^T V = \rh W^{-1} \sigma_3 W\rh^{-1}\ ,\qquad \eta = \begin{pmatrix}  0 & {\mathbb I} \\ {\mathbb I}  & 0 \end{pmatrix}\ ,  \qquad \sigma_3 = \begin{pmatrix} {\mathbb I} & 0 \\ 0 & -{\mathbb I} \end{pmatrix}\ .
\ee
It follows that
\be
\partial_\m {\cal S} =\rho W^{-1} \big[ W \pd_\m W^{-1} , \sigma_3 \big] W \rho^{-1} = 
\rho W^{-1}  \left( \begin{array}{cc} 0	&	2 P_\m	\\ - 2 P_\m^T	&	0
\end{array} \right)  W \rho^{-1}\ .
\ee
Thus we derive the identities,
%^Q
\begin{align}
\tr ( \pd_\m \mathcal{S}\, \pd_\n \mathcal{S} ) =&\, - 4 \tr ( P_\m P_\n^T ) - 4 \tr ( P_\m^T P_\n ) \ ,
\nn\w2
\tr ( \pd_\m \mathcal{S}\, \pd_\n \mathcal{S} ) \tr ( \pd^\m \mathcal{S}\, \pd^\n \mathcal{S} ) =&\, 64 \tr( P_\m P_\n^T ) \tr ( P^\m P^{\n T} )\ , 
\nn\w2
\tr ( \pd_\m \mathcal{S}\, \pd^\m \mathcal{S}\, \pd_\n \mathcal{S}\, \pd^\n \mathcal{S} ) =&\, 16 \tr ( P_\m P^{\m T} P_\n P^{\n T} ) + 16 \tr ( P_\m^T P^\m P_\n^T P^\n ) \ ,
\nn\w2
\tr ( \pd_\m \mathcal{S}\, \pd_\n \mathcal{S}\, \pd^\m \mathcal{S}\, \pd^\n \mathcal{S} ) =&\, 32 \tr ( P_\m P_\n^T P^\m P^{\n T} ) \ ,
\nn\w2
\tr ( \mathcal{S}\, \pd_\m \mathcal{S}\, \pd^\m \mathcal{S}\, \pd_\n \mathcal{S}\, \pd^\n \mathcal{S} ) =&\, 16 \tr ( P_\m P^{\m T} P_\n P^{\n T} ) - 16 \tr ( P_\m^T P^\m P_\n^T P^\n ) \ .
\end{align}
Using these identities, \eq{LB2} takes the form 
\begin{align} 
\cL_{Bos. , \OA} =&  e e^{2\varphi}
\Bigg\{ H^{\m\n\rh} \big( \omega^L_{\m \n \rh} (\omega) +\omega^Q_{\m\n\rh}(Q_-) \big)
 - \tfrac14 R_{\mu\nu mn}(\omega) R^{\mu\nu mn}(\omega) 
\nn\w2
& +\tfrac12 R_{\m\n\rh\s} H^{\m\n,\rh\s}  +\tfrac12 H^2_{\m\n} H^{2 \m\n} - \tfrac16 H_{\m\n,\rh\s} H^{\m\rh,\n\s} - \tfrac18 H^{2\m\n} \tr (\partial_\m {\cal S} \partial_\n {\cal S}) 
\nn\w2
& + \tfrac1{32} \Big[ \tfrac14 \tr ( \pd_\m \mathcal{S}\, \pd_\n \mathcal{S} ) \tr ( \pd^\m \mathcal{S}\, \pd^\n \mathcal{S} ) - \tfrac12 \tr ( \pd_\m \mathcal{S}\, \pd_\n \mathcal{S}\, \pd^\m \mathcal{S}\, \pd^\n \mathcal{S} )
\nn\w2
& - \tr ( \mathcal{S}\, \pd_\m \mathcal{S}\, \pd^\m \mathcal{S}\, \pd_\n \mathcal{S}\, \pd^\n \mathcal{S} ) \Big]\Bigg\}\ .
\label{LB4}
\end{align}
Finally, we note that the CS form satisfies
\be
d \Big[ \omega^Q(Q_+) +\omega^Q(Q_-)\Big]=0\ ,
\ee
which implies
\be
\omega^Q(Q_+)= -\omega^Q(Q_-) + d\theta\ ,
\ee
for some 2-form $\theta$. With this relation at hand, we find that our result \eq{LB4} agrees with that of \cite{Eloy:2020dko} in their eq. (7.16), upon setting the vector fields equal to zero, and taking into account the convention differences. Similarly we also find that our results agree with those of \cite{Baron:2017dvb}\footnote{We thank Carmen Nunez for communications on this comparison.}.
On the other hand, our results apparently differ from those given in \cite{Ortin:2020xdm}.

%%%%%%%%%%%%%%%%%%%%%%%%%%
\section{Conclusions}
%%%%%%%%%%%%%%%%%%%%%%%%%%

 Motivated by the exploration of higher derivative couplings of quaternionic Kahler sigma models to $N=(1,0)$ supergravity in $6D$, we have started with heterotic supergravity at $\OA$ \cite{Bergshoeff:1989de}, and reduced it on $T^4$ with a consistent $N=(1,0)$ supersymmetric truncation. 
 We have found that the manifest rigid  $GL(4)$ and composite local $SO(4)$ symmetry gets enhanced to rigid $SO(4,4)$ and composite local $SO(4)_+\times SO(4)_-$, with the hyperscalars parametrizing the Grassmannian coset $Gr(4,4)$.
 A series of field redefinitions in the hypermultiplet sector are found to cancel a large number of terms arising in the reduction of the action and supersymmetry transformation rule that have only $O(4)$ invariance. These results generalize the well known work of Maharana and Schwarz \cite{Maharana:1992my} who showed how the $O(d,d)$ invariance emerges in the bosonic action and at the two-derivative level, and the results of \cite{Eloy:2020dko,Ortin:2020xdm} where the $\OA$ terms in the bosonic action were dimensionally reduced. We have also shown that the treatment of the 3-form field strength in heterotic supergravity as torsion part of the spin connection, and the modification of its field strength by Lorentz Chern-Simons form defined in terms of the torsionful spin connection, simplify the reduction considerably. In the resulting $6D$ Lagrangian, many $H$ dependent terms are absorbed into a torsionful spin connection, but there exist terms in which the three form field strength appears explicitly.

The cancellations of duality symmetry offending terms is expected in view of Sen's result based on string field theory \cite{Sen:1991zi}. However, the emergence of the duality symmetry at the field theory level is a nontrivial symmetry enhancement phenomenon, which remains to be better understood. The requirement that the dimensionally reduced supersymmetry transformations take an appropriate form may provide a good start for understanding the field redefinitions in a simpler way. The inclusion of the abelian sector of the Yang-Mills couplings in heterotic supergravity remains to be carried out, and it is expected to give the $O(20,4)$ symmetry in $6D$. 

One of the motivations for the current work has been the construction of higher derivative couplings of $N=(1,0)$ supergravity to hypermultiplets where the hyperscalars parametrize a noncompact quaternionic Kahler sigma model with negative curvature constant. The Grassmannian coset $Gr(n,4)$ is one of the Wolf spaces that have this property. For $n=4$ we have shown explicitly here how this coupling emerges from dimensional reduction. The complexity of the result shows that a direct construction of these couplings by means of Noether procedure would be 
very complicated. Dimensional reduction proves a relatively simpler approach to this problem. However, there are no compactification schemes that we know of for obtaining the higher derivative couplings of the other QK sigma models that are relevant to $6D$ supergravity couplings. For those cases, apparently we need to resort to the Noether procedure. The results of the current paper provide a guide in writing down an ansatz for the all possible four-derivative couplings in this construction. The consequences of the fact that the structure group in $Gr(n,4)$ is $SO(n)\times SO(3) \times Sp(1)_R$, while in the other Wolf spaces it is either $SU(n)\times U(1)\times Sp(1)_R$, or $G\times Sp(1)_R$ where $G$ is a particular simple group, remains to be investigated. Ultimately, 
the Yang-Mills sector is to be included, and $R$-symmetry is to be gauged, in an anomaly-free fashion. The investigation of dyonic string solutions in that framework is expected to play role in the analysis of the consistencies of these theories \cite{Pang:2020rir}.

\subsubsection*{Acknowledgements}

Special thanks are due to Guillaume Bossard for helpful discussions and for suggesting the dimensional reduction of higher derivative heterotic supergravity. We also thank Katrin Becker, Eric Bergshoeff, Daniel Butter, Falk Hassler, Olaf Hohm, Axel Kleinschmidt, James Liu, Yi Pang and Toine van Proeyen for useful discussions, and  Jose Fernandez-Melgarejo and Carmen Nunez for correspondence on the comparison of our results. The work of ES is supported in part by the NSF grants PHY-1803875 and PHYS-2112859, and the work of HC is supported in part by the Mitchel Institute for Physics and Astronomy.

\begin{appendix}

%%%%%%%%%%%%%%%%%%%%%%%%%%%%%%%%%%%%%%%%%
\section{Notation and conventions}
%%%%%%%%%%%%%%%%%%%%%%%%%%%%%%%%%%%%%%%%%

In our conventions the spacetime signature is $(-++...+)$, and the fields of \cite{Bergshoeff:1989de} need to be scaled as follows: 
\bea
&& \psi_\m \to \sqrt 2 \psi_\m\ ,\quad \ \e \to \sqrt 2 \e , \quad \ B_{\m\n}\to -\sqrt 2 B_{\m\n}\ ,
\quad H_{\m\n\rh} \to -(\sqrt 2/3) H_{\m\n\rh}\ , 
\nn\w2
&&  \phi \to \exp (-2\vp/3)\ ,\quad \omega \to -\omega\ ,\quad   \Omega_\pm \to -\Omega_\pm\ .
\eea
We frequently use the definitions
\begin{align}
X_{\m\n ab} :=& P_{\m a}{}^c P_{\n cb}\ , & Y_{\m\n ab}:=& P_{\m a}{}^c P_{\n bc}\ , & Z_{\m\n ab} :=& P_{\m c a} P_\n{}^c{}_b\ ,
\nn\w2
X_{\m\n} := & \delta^{ab} X_{\m\n ab}\ , & Y_{\m\n} :=& \delta^{ab} Y_{\m\n ab} \ , &  Y_{\m\n} =& \delta^{ab} Z_{\m\n ab}\ ,
\nn\w2
X_{ab} := & g^{\m\n} X_{\m\n ab}\ ,& Y_{ab} :=&  g^{\m\n}  Y_{\m\n ab}\ , & Z_{ab} :=& g^{\m\n}  Z_{\m\n ab}
\ .
\label{def3}
\end{align}
Thus, $Y_{\m\n [ab]}= -\tfrac12 Q_{+\m\n ab}$ and $Z_{\m\n [ab]}= -\tfrac12 Q_{-\m\n ab}$. The vielbein postulates are
\begin{equation}
\begin{aligned}
\pd_\m e_\n^m + \om_\m{}^m{}_n e_\n{}^n - \Gamma_{\m\n}{}^\ta e_\ta{}^m =&\, 0 \ , 
\\
\pd_\m e_\n^m + \Omega_{\pm \m}{}^m{}_n e_\n{}^n - \Gamma_{\mp \m\n}{}^\ta e_\ta{}^m =&\, 0 \ , 
\label{vp}
\end{aligned}
\end{equation}
where 
\be
\Omega_{\pm \m rs} = \omega_{\m rs} \pm H_{\m rs}\ ,\qquad \Gamma_{\pm\m\n}^\rh = \Gamma_{\m\n}^\rh \pm H_{\m\n}{}^\rh\ ,\qquad H_{\m\n\rh} = 3\partial_{[\m} B_{\n\rh]}\ ,
\ee
and $\Gamma_{\m\n}^\rh$ represents the torsion-free Christoffel symbol. The gamma matrices are covariantly constant,
\begin{equation}
D_\m(\om, \Gamma) \g_\n = 0 \ , \qquad D_\m(\Omega_\pm, \Gamma_\mp) \g_\n = 0 \ . 
\label{gmc}
\end{equation}
The curvatures are defined as
\begin{align}
R_{\m\n}{}^{mn}(\om) \equiv&\, \pd_\m \om_\n{}^{mn} - \pd_\n \om_\m{}^{mn} + \om_\m{}^{mp} \om_{\n p}{}^n - \om_\n{}^{mp} \om_{\m p}{}^n\ ,
\label{a3}
\w2
R_{\m\n}{}^\rh{}_\s(\Gamma) \equiv&\, \pd_\m \Gamma_{\n\s}{}^\rh - \pd_\n \Gamma_{\m\s}{}^\rh + \Gamma_{\m\ta}{}^\rh \Gamma_{\n\s}{}^\ta - \Gamma_{\n\ta}{}^\rh \Gamma_{\m\s}{}^\ta\ .
\label{a4}
\end{align}
%
%Notice that \eqref{a4} is different from (A.1) of \cite{Bergshoeff:1986wc}. 
The two curvatures are related by
\be
R_{\m\n\rh\s}(\Gamma) = R_{\m\n rs}(\om) e_\rh{}^r e_\s{}^s \ , \qquad R_{\m\n\rh\s}(\Gamma_\pm) = R_{\m\n mn}(\Omega_\mp) e_\rh{}^m e_\s{}^n \ . 
\label{a5}
\ee
These identities can be easily derived by considering commutator of covariant derivatives acting on vielbein, and make use of \eqref{vp}. The curvatures of $\Gamma_\pm$ are related to $\Gamma$ by
\begin{equation}
R_{\m\n\rh\s}(\Gamma_\pm) = R_{\m\n\rh\s}(\Gamma) \mp 2 D_{[\m}(\Gamma) H_{\n]\rh\s} + 2 H_{[\m|\rh}{}^\ta H_{|\n] \ta \s} \ . 
\label{a6}
\end{equation}
The curvatures of $\Gamma_\pm$ are related to each other by
\begin{equation}
R_{\m\n\rh\s}(\Gamma_+) = R_{\rh\s\m\n}(\Gamma_-) \ . 
\label{a8}
\end{equation}
We also have the relation
\bea
D_\mu(\Omega_-) P_{m ab} 
&:=& \partial_\mu P_{mab} + \Omega_{-\mu m}{}^n P_{nab} 
 + Q_{+\mu a}{}^c P_{mcb} + Q_{-\mu b}{}^c P_{mac}
 \nn\w2 
 &=& e_m{}^\n D_\m (\Gamma_+) P_{\n ab}\ .
\eea
with $D_\mu(\Gamma_+) P_{\n ab}$ as defined in \eq{DP2}, and 
\be
D_{[\m} (\Gamma_+) P_{\n] ab} = - H_{\m\n}{}^\rh\,P_{\rh ab}\ .
\ee
Finally, our notation for the covariant derivatives is as follows. From Section 4 onward, in covariant derivatives we only indicate the connections that act on the Lorentz spinor and vector indices, and suppress the composite local connections $Q_\pm$ that act  according to the $SO(4)_+\times SO(4)_-$ representations carried by the fields they act on. When we encounter a term in which this symmetry is broken, we display the composite connections in the covariant derivatives. For reader's convenience, we list the definition of variety of covariant derivatives that arise in the body of the paper:
\allowdisplaybreaks{
\begin{align}
D_{\mu}(\Omega_+) \epsilon
=& \left( \partial_{\mu} + \tfrac{1}{4} \Omega_{+\mu mn} \gamma^{mn} 
+ \tfrac{1}{4} Q_{+\mu ab} \Gamma^{ab} \right) \epsilon\ ,
\nn\w2
D_{[\mu}(\omega) \psi_{\nu]} 
=& \left( \partial_{[\mu|} + \tfrac{1}{4} \omega_{[\mu| rs} \gamma^{rs} 
+ \tfrac{1}{4} Q_{+[\mu| ab} \Gamma^{ab} \right) \psi_{|\nu]}\ , 
\nn\w2
D_{\mu}(\omega) \chi
=& \left( \partial_{\mu} + \tfrac{1}{4} \omega_{\mu rs} \gamma^{rs} 
+ \tfrac{1}{4} Q_{+\mu ab} \Gamma^{ab} \right) \chi\ , 
\nn\w2
D_{\mu}(\omega) \psi_a
=& \left( \partial_{\mu} + \tfrac{1}{4} \omega_{\mu rs} \gamma^{rs} 
+ \tfrac{1}{4} Q_{+\mu cd} \Gamma^{cd} \right) \psi_a
+ Q_{-\mu a}{}^b \psi_b\ ,
\nn\w2
D_\m (\omega,\Omega_-) \psi_r =&  \left(\partial_\m +\tfrac14 \omega_{\m pq} \gamma^{pq} +\tfrac14 Q_{+\m ab} \Gamma^{ab} \right)\psi_r + \Omega_{-\m r}{}^s \psi_s\ ,
\nn\w2
D_\m \psi_a =& D_\m (\Omega_+)\psi_a = \left( \partial_{\mu} + \tfrac{1}{4} \Omega_{+\mu rs} \gamma^{rs} 
+ \tfrac{1}{4} Q_{+\mu cd} \Gamma^{cd} \right) \psi_a + Q_{-\mu a}{}^b \psi_b\ ,
\nn\w2
D_\m P_{\n ab} =& D_\m(\Gamma_+) P_{\n ab} =\partial_\m P_{\n ab} -\Gamma_{+\m\n}{}^\rh P_{\rh ab}  + Q_{+\m a}{}^c P_{\n cb} + Q_{-\mu b}{}^c P_{\n ac}\ ,
\nn\w2
D_\mu(\omega, \Omega_-) \psi_{rs} 
=& \left( \partial_{\mu} + \tfrac{1}{4} \omega_{\mu pq} \gamma^{pq} 
+ \tfrac{1}{4} Q_{+\mu ab} \Gamma^{ab}\right) \psi_{rs}
+ \Omega_{-\mu r}{}^p \psi_{ps} + \Omega_{-\mu s}{}^p \psi_{rp}\ .
\label{CDs}
\end{align}}
Further definitions are
\begin{align}
\cH_{\m\n\rh} =& 3\partial_{[\m} B_{\n\rh]}  -6\alpha' \tr \left(
\Omega^{(sc)}_{-[\m} \partial_\n \Omega^{(sc)}_{-\rh]} + \tfrac23 \Omega^{(sc)}_{-[\m}\Omega^{(sc)}_{-\n} 
\Omega^{(sc)}_{-\rh]} \right) 
\nn\w2
& -6\alpha'\tr \left( Q^{(sc)}_{-[\mu} \partial_\nu Q^{(sc)}_{-\rho]} 
+ \tfrac23 Q^{(sc)}_{-[\mu} Q^{(sc)}_{-\nu} Q^{(sc)}_{-\rho]} \right)\ ,
\nn\w2
\Omega^{(sc)}_{-\mu rs} =& \Omega_{-\mu rs} + \bar{\psi}_r \gamma_\mu \psi_s\ , 
\nn\w2
Q^{(sc)}_{-\mu ab} =& Q_{-\mu ab} + \bar{\psi}_a \gamma_\mu \psi_b\ ,
\nn\w2
D_\m^{(sc)} \psi_a =& D_\m \psi_a +\tfrac12 \gamma^\nu\Gamma^b \psi_\m P_{\n ba} \ ,
\nn\w2
P_{\m ab}^{(sc)} =& P_{\m ab} -\bpsi_\m \Gamma_a \psi_b\ ,
\nn\w2
Y_{\m\n}^{(sc)} =& P_{\m ab}^{(sc)} P_\n{}^{(sc)\,ab}\ ,
\nn\w2
Q_{-\m\n ab}^{(sc)} =& -2 P_{[\m}{}^{ca (sc)} P_{\n]c}{}^{b (sc)}\ ,
\nn\w2
Z_{ab}^{(sc)} =& P_{\m ca}^{(sc)} P^{\m c}{}_b{}^{(sc)}\ ,
\label{mdefs}
\end{align}
where terms up to quadratic in fermions are to be kept in the last three equations.

%%%%%%%%%%%%%%%%%%%%%%%%%%%%%%%%%
\section{Field redefinitions}
%%%%%%%%%%%%%%%%%%%%%%%%%%%%%%%%%

Consider the field redefinition $E_\alpha{}^a \to E'_\alpha{}^a + \delta E_\al{}^a$ with
\begin{equation}
\delta E_\al{}^a = E_\al{}^b S_b{}^a \ ,  
\end{equation}
where $S_{ab} = S_{ba}$.
Under this redefinition, 
\begin{align}
\delta Q_{\pm \m a b} =&  P_{\mp \m a}{}^c S_{cb} - P_{\mp \m b}{}^c S_{ac} \ ,
\nn\w2
\delta P_{\pm \m a b} =& \pd_\m S_{ab} + Q_{\mp \m a}{}^c S_{cb} + Q_{\pm \m b}{}^c S_{ac} \ ,
\nn\w2
\delta P_{\m a b} =&  D_\m(Q_+) S_{ab} +\big( P_{\m b}{}^c -P_\m{}^c{}_b \big) S_{ac} \ .
\label{dpq}
\end{align}
The variation of the zeroth order Lagrangian under this redefinition is 
\begin{align}
\delta \mathcal{L}_0 &= e e^{2\varphi} \biggl[ 
\Big( \tfrac18 \bar{\psi}_\n \gamma^{\mu\nu\rho}   \Gamma^{ab} \psi_\rho 
+  \tfrac12  \bar{\chi} \gamma^{\mu\nu} 
\Gamma^{ab} \psi_\nu 
+ \tfrac12 \bar{\chi} \gamma^\mu  \Gamma^{ab} \chi -\tfrac18  \bar{\psi}^c \gamma^\mu  \Gamma^{ab}\psi_c \Big) \delta Q_{+\m ab}
\nn\w2
&-\tfrac12 \bpsi^a \gamma^\m \psi^b \delta Q_{-\mu ab}
+ \tfrac12 \left( 
\bar{\psi}_\nu \gamma^\mu \gamma^\nu \Gamma^a \psi^b + 2 \bar{\chi} \gamma^\mu \Gamma^a \psi^b \right) \delta P_{\mu ab}  - \tfrac12 P^{\mu ab} \delta P_{\mu ab} 
\biggr]\ .
\label{YT1}
\end{align} 
Let us now consider the redefinition
\begin{align}
S^{ab} =& -2\alpha'\, Y^{ab} \ .
\label{pr1}
\end{align}
It gives rise to 
\begin{align}
\delta \cL_0 \Big|_{E \to YE} =&\, \al'\, e\, e^{2 \vp} \Bigg[ P^{\m ab} D_\m Y_{ab} +X_{ab} Y^{ab} -Z_{\m\n ab}Z^{\m\n ab} 
\nn\w2
&  -\Big( \tfrac12 \bpsi_\n \g^{\m\n\rh} \Gamma^{ab} \psi_\rh +2 \bar{\chi} \g^{\m\n} \Gamma^{ab} \psi_\n + 2 \bar{\chi} \g^\m \Gamma^{ab} \chi 
\nn\w2
& - \tfrac12 \bpsi^d \g^\m \Gamma^{ab} \psi_d  \Big) ( P_{\m a}{}^c Y_{bc} )
+2\bpsi^a \g^\m \psi^b \big(P_\m{}^c{}_a Y_{bc} \big)
\nn\w2
&   + \Big(- \bpsi_\n \g^\m \g^\n \Gamma^a \psi^b - 2 \bar{\chi} \g^\m \Gamma^a \psi^b \Big) ( P_{\m b}{}^c Y_{ac})
\nn\w2
& - \Big( \bpsi_\n \g^\m \g^\n \Gamma^a \psi^b + 2 \bar{\chi} \g^\m \Gamma^a \psi^b \Big) D_\m Y_{ab} 
\nn\w2
& +\Big(\bpsi_\n \g^\m \g^\n \Gamma^a \psi^b + 2 \bar{\chi} \g^\m \Gamma^a \psi^b \Big) P_\m{}^c{}_b Y_{ac} \Bigg]\ . 
\label{SL}
\end{align}
The term $-e e^{2\vp} Z_{\m\n ab} Z^{\m\n ab}$ and the last two terms are $SO(4)_+ \times SO(4)_-$ invariant. The rest will remove some of the terms that break this symmetry as shown in \eq{ML1}. The remaining symmetry breaking terms will be removed by further field redefinitions discussed below. Next, let us consider the redefinition
\be
S_{ab} = 4\alpha'\, \bpsi^\n \Gamma^{(a} \psi_c P_\n{}^{b)c}\ .
\label{pr2}
\ee
In this case only the last term in \eq{YT1} contributes, giving
\begin{align}
\delta \cL_0 \Big|_{S-\rm terms} =& -\tfrac12 e e^{2\varphi} \Big[ P^{\m ab} D_\m (Q_+,Q_-) S_{ab} \Big]\ ,
\label{ML}
\end{align}
where $Q_+$ rotates the first index and $Q_-$ rotates the second index on $S_{ab}$. Integration by part gives
\be
\delta \cL_0 \Big|_{E\to \psi_\n \psi_a PE} = 2 e\bpsi^\n\Gamma^{(a}\psi_c P_\n{}^{b)c} D_\m\big(e^{2\vp} P^\m{}_{ab}\big)\ .
\ee
Next, we consider the redefinition of the hyperino. The lowest order Lagrangian $\cL_0$ under a general variation of the hyperino gives
\begin{align}
\delta \cL_0 =&\, e\, e^{2 \vp} \Bigg[ - \tfrac12 ( \delta \bpsi^a ) \slashed{D}(\omega) \psi_a - \tfrac12 \bpsi^a \slashed{D}(\omega)( \delta \psi_a ) 
\nn\w2
& - \tfrac{1}{12} H_{\m\n\rh} \bpsi^a \g^{\m\n\rh} \delta \psi_a + \tfrac12 P_{\m a b} \Big( \bpsi_\n \g^\m \g^\n \Gamma^a \delta \psi^b + 2 \bar{\chi} \g^\m \Gamma^a \delta \psi^b \Big) \Bigg] \ . 
\label{gv2}
\end{align} 

It follows that the field redefinition 
\begin{equation}
\delta \psi_a = -2 \alpha' \psi^b Y_{ab} - 2 \alpha' \g^\n \Gamma^b \psi^\m Y_{\m\n ab} -4 \alpha' P_{\m ab} D^\m \psi^b \ ,
\end{equation}
yields the results 
\begin{align}
\delta \cL_0 \Big|_{\psi \to \psi Y} =&\, \alpha'\, e\, e^{2 \vp} \Big[ 2 Y_{ab}\, \bpsi^a \slashed{D}(\om) \psi^b + \tfrac16 H_{\m\n\rh} \bpsi^a \g^{\m\n\rh} \psi^b Y_{ab} 
\nn\w2
& - ( \bpsi_\n \g^\m \g^\n \Gamma^a \psi^b + 2 \bar{\chi} \g^\m \Gamma^a \psi^b ) ( P_{\m a}{}^c Y_{bc} ) \Big] \ , \end{align}
\begin{align}
\delta \cL_0 \Big|_{\psi^a \to \psi^\m PP} =&\, 
 \alpha'\, e\, e^{2 \vp} \Big[ Y_{\m\n ab}\,\bpsi^b \g^\rh \g^\m \Gamma^a  D_\rh(\omega, \Gamma_+) \psi^\n + Y_{\m\n ab}\,\bpsi^\m \Gamma^b \g^\n \slashed{D}(\omega) \psi^a   
\nn\w2
&  - \bpsi^\m \g^\n \g^\rh \Gamma^a \psi^b D_\rh Y_{\m\n ba} 
 + \tfrac16 H_{\m\n\rh} \bpsi^a \g^{\m\n\rh} \g^\s \Gamma^b \psi^\ta Y_{\ta\s ab}
\nn\w2
& + \bpsi_\rh \g^\m \g^\rh \g^\n \Gamma^a \Gamma^b \psi^\s ( P_{\m a c} Y_{\n\s bc} ) + 2 \bar{\chi} \g^\m \g^\n \Gamma^b \Gamma^d \psi^\s ( P_{\m b c} Y_{\n\s dc} ) 
\nn\w2
& - \bpsi^\m \g^\n \g^\rh \Gamma^a \psi^b P_{\rh b}{}^c Y_{\m\n c a} + \bpsi^a \g^{\n\rh} \Gamma^b \psi^\m H_{\n\rh}{}^\ta Y_{\m\ta a b} 
\nn\w2
& - \bpsi^a \gamma^\rh \gamma^\m \Gamma^b \psi^\n \big( P_\rh{}^c{}_a Y_{\m\n bc} \big) \Big]\ , 
\label{rd2}
%%%%%%%%%%%
\w4
\delta \cL_0 \Big|_{\psi \to P D \psi} =&\, \alpha'\, e\, e^{2 \vp} \Bigg[ 2 P_{\m a b} \bpsi^a \slashed{D}(\omega, \Gamma_+)  D^\m \psi^b  + 2 ( \bpsi^b \g^\m D^\n \psi^a ) D_\m P_{\n b a}
\nn\w2
& + 2P_{m b a} ( D^m \bpsi^a ) \slashed{D}(\omega) \psi^b   + \tfrac13 H_{\m\n\rh} \bpsi^a \g^{\m\n\rh} ( D_m \psi^b ) P^m{}_{ab} 
\nn\w2
& - 2 \bpsi^\rh \g^\m \g_\rh \Gamma^b ( D^\n \psi^a ) P_{\m b}{}^c P_{\n c a} - 4 \bar{\chi} \g^\m \Gamma^b ( D^\n \psi^a ) P_{\m b}{}^c P_{\n c a} 
\nn\w2
& + 2 \bpsi^a \g^\m ( D^\n \psi^b ) X_{\m\n a b}  - 2 \bpsi^a \g^\m ( D^\n \psi^b ) Z_{\m\n a b}\Bigg]\ , 
\label{rd3}
\end{align}
where $\Gamma$ refers to the Christoffel symbol which is torsion-free. The last term in \eq{rd2} and the last term in \eq{rd3} are $SO(4)_+ \times SO(4)_-$ invariant. The rest will remove the remaining symmetry breaking terms, as shown in \eq{ML1}. In the above equations we have used the following notations. 
We have denoted the torsionful connection by $\Gamma_\pm = \Gamma \pm H$. In \eq{rd2} we have converted $D_\rh (\Gamma,Q_-,Q_+) Y_{\m\n ba}$ in which the connection $Q_{-}$ acts on the $b$ and $Q_{+}$ acts on $a$ index, to the standard one $D_\rh (\Gamma_+,Q_+,Q_+) Y_{\m\n ba} = D_\rh Y_{\m\n ba}$ by adding and subtracting the required terms. In \eq{rd2} we have also converted $D_\rh(\omega,\Gamma)\psi_\n$ where $\omega$ rotates the spinor index and $\Gamma$ acts on the vector index of the gravitino, to $D_\rho(\omega,\Gamma_+) \psi_\n$, again by adding and subtracting the required terms. Similarly, in \eq{rd3}, we have converted $ \slashed{D}(\omega) D^\m \psi^b$ into $\slashed{D}(\omega, \Gamma_+) D^\m \psi^b$, and $D_\m(\Gamma,Q_-,Q_-) P_{\n ba}$ into $D_\mu(\Gamma_+,Q_+,Q_-) P_{\nu ba} = D_\m P_{\n ba}$, again by adding and subtracting appropriate terms.

%%%%%%%%%%%%%%%%%%%%%%%%%%%%%%%%%%%%%%
\section{The total Lagrangian in $6D$ }
%%%%%%%%%%%%%%%%%%%%%%%%%%%%%%%%%%%%%%

A number of terms in the Lagrangian \eq{ML1} can be absorbed to the definition of $H$ extended to $\cH$ defined in \eq{mdefs} in the lowest order Lagrangian \eq{6D}, which we reproduce here for reader's convenience
\begin{align}
\mathcal{L}_0 &= e e^{2\varphi} \Big[ \, \tfrac{1}{4} R 
+ g^{\mu\nu} \partial_\mu \varphi \partial_\nu \varphi 
- \tfrac{1}{12} H_{\mu\nu\rho} H^{\mu\nu\rho} - \tfrac{1}{4} P_{\mu ab} P^{\mu ab}
\nn\\ &\quad 
- \tfrac{1}{2} \bar{\psi}_\mu \gamma^{\mu\nu\rho} 
D_\nu(\omega) \psi_\rho 
+ 2 \bar{\chi} \gamma^{\mu\nu} D_\mu(\omega) \psi_\nu 
+ 2 \bar{\chi} \gamma^\mu D_\mu(\omega) \chi 
\nn\\&\quad
 - \tfrac{1}{2} \bar{\psi}^a \gamma^\mu D_\mu(\omega) \psi_a  - \partial_\mu \varphi \left( 
\bar{\psi}^\mu \gamma^\nu \psi_\nu 
 + 2 \bar{\psi}_\nu \gamma^\mu \gamma^\nu \chi \right) 
\nn\\&\quad
+ \tfrac{1}{2} P_{\mu ab} \Big( 
\bar{\psi}_\nu \gamma^\mu \gamma^\nu \Gamma^a \psi^b 
+ 2 \bar{\chi} \gamma^\mu \Gamma^a \psi^b \Big) 
- \tfrac{1}{24} H_{\mu\nu\rho} \Big( 
\bar{\psi}^\sigma \gamma_{[\sigma} \gamma^{\mu\nu\rho} 
\gamma_{\tau]} \psi^\tau 
\nn\\&\quad 
+ 4 \bar{\psi}_\sigma \gamma^{\sigma\mu\nu\rho} \chi 
- 4 \bar{\chi} \gamma^{\mu\nu\rho} \chi +  \bar{\psi}^a \gamma^{\mu\nu\rho} \psi_a   \Big) 
\Big]\ .
\label{6D2}
\end{align}
In addition to the replacement $H$ by $\cH$ in $\cL_0$, we can also use of the supercovariantized torsionful connection $\Omega_-^{(sc)}$ in the ${\rm Riem}^2(\Omega_-)$ term. In doing so, the following relations are useful:
\begin{align}
-\tfrac1{12} \cH_{\m\n\rh} \cH^{\m\n\rh} =&\,   -\tfrac1{12}  
H_{\m\n\rh} H^{\m\n\rh} +\alpha' H^{\m\n\rh} \big( \omega^L_{\m\n\rh} (\Omega_-) + \omega^Q_{\m\n\rh}(Q_-) \big) 
\nn\w2
& - \alpha' H^{\m\n\rh} R_{\m\n}{}^{rs}(\Omega_-)\,\bpsi_r \gamma_\rh \psi_s - \alpha' H^{\m\n\rh} Q_{-\m\n}{}^{ab}\,\bpsi_a \gamma_\rh \psi_b 
\nn\w2
& + \alpha' \big( \bpsi_r\gamma_\n\psi_s \Omega_{-\rh}{}^{rs} 
+ \bpsi_a\gamma_\n\psi_b Q_{-\rh}{}^{ab}\big) e^{-2\vp} D_\m(\Gamma)\big(e^{2\vp} H^{\m\n\rh}\big) \ ,
\nn\w2
 -\tfrac14 \alpha' R_{\mu\nu rs}(\Omega_-^{(sc)}) R^{\mu\nu rs}(\Omega_-^{(sc)} ) =&\, -\tfrac14 \alpha' R_{\mu\nu rs}(\Omega_-) R^{\mu\nu rs}(\Omega_-) 
\nn\w2
& -2\alpha' R_{\m\n rs}(\Omega_-)   \bpsi^r \gamma^\n D^\m (\omega,\Omega_-)\psi^s\ ,
\end{align}
where $H=dB$. Furthermore, carrying out the algebra of Dirac matrices to determine the independent structures, and to separate terms are amenable to the use of the lowest order field equations, remarkable simplifications occurs and the total $6D$ Lagrangian takes the form
%
%%%%%%%%%%%%%%%%%    ORGANIZED RESULT   %%%%%%%%%%%%%%%%%
%
\begin{align}
\cL = & \cL_0\Big|_{H\to \cH } +\cL(R^2) + \cL_1 + \cL_2 +\cL_3+\cL_4+\cL_5+\cL_6\ ,
\label{OrgL}
\end{align}
with $\cL_0$ as given in \eq{6D}, and various parts of the Lagrangian are organized according to the structures of the terms they consist of as follows: 
\allowdisplaybreaks{
\begin{align}
\cL(R^2) =& \alpha' e e^{2\varphi}\Big[ 
- \tfrac14 R_{\mu\nu rs}(\Omega_-^{(sc)}) R^{\mu\nu rs}(\Omega_-^{(sc)})
 - \bar{\psi}^{rs} \gamma^\mu D_\mu (\omega,\Omega_-)\,  \psi_{rs}
- \tfrac{1}{12} H_{\mu\nu\rho} 
\bar{\psi}^{rs} \gamma^{\mu\nu\rho} \psi_{rs}
\nn\w2
& \qquad\quad + \tfrac12 R_{\m\n}{}^{rs} (\Omega_-) \big(\bpsi^\rh \gamma^{\m\n}\gamma_\rh -2{\bar\chi}\gamma^{\m\n} \big) \psi_{rs} \Big]
\w4
\cL_1 =& \alpha' e e^{2\varphi}\Big[ - \big(D_\mu P^{(sc)}_{\n ab}\big) D^\m P^{\n ab}_{(sc)} -\tfrac34  Q_{-\mu\nu ab}^{(sc)} Q_-{}^{(sc) \mu\nu ab}
-\tfrac12 Y_{\m\n}^{(sc)} Y^{(sc)\,\m\n} 
\nn\w2
& \qquad\quad -\tfrac12 Z_{ab}^{(sc)}Z^{(sc) ab} 
 \Big]
\w4
\cL_2 =&  \alpha' e e^{2\varphi}\Big[ \big( -\tfrac12\bpsi_\n \gamma^{\m\n\rh} \Gamma^{ab} \psi_\rh 
-2\bpsi_\n \gamma^{\m\n} \Gamma^{ab} \chi -2{\bar\chi} \gamma^\m \Gamma^{ab} \chi 
+ \tfrac12\bpsi^d\gamma^\m \Gamma^{ab}  \psi_d \big) P^\s{}_a{}^c D_\m P_{\s bc}  
\nn\w2
& \qquad \quad-2\bpsi_\rh\gamma_\m\psi_\n \big(P^\n{}_{ab} D^\m P^{\rh ab}\big)
-2\bpsi_b\gamma^\m\psi_c \big(P^{\n ac}  D_\m P_{\n a}{}^b\big) \Big]
\w4
\cL_3 =&  \alpha' e e^{2\varphi}\Big[-\tfrac12 \bpsi_\s \gamma^\s \gamma^{\m\n\rh} \Gamma^a \psi^b P_\m{}^c{}_b Y_{\n\rh ca} -\tfrac14 \bpsi_\m \gamma^\m \gamma^\n \Gamma^{cda} \psi^b P^\rh{}_{db} Q_{+\n\rh ac} 
\nn\w2
& \qquad\quad -\tfrac12 \bpsi_\m \gamma^\m \gamma^\n \Gamma^a \psi^b \big( P^\rh{}_{ab} Y_{\n\rh} -2P^{\rh c}{}_b Y_{\n\rh ca} +3 P_\n{}^c{}_b Y_{ca}   \big)
\nn\w2
&  \qquad\quad -{\bar\chi}\gamma^{\m\n}\gamma^\s \Gamma^c\psi_a\, 
P_{\s cb} Q_{-\m\n}{}^{ab}  -2{\bar\chi}\Gamma^{cd} \Gamma^b \gamma^\nu \psi^a\, P^\m{}_{ca} Y_{\m\n db}
\nn\w2
& \qquad\quad  + 2 \bar{\chi} \g^\m \Gamma^a \psi^b  P_\m{}^c{}_b Y_{ac} \Big]
\w4
\cL_4 =& \alpha' e e^{2\varphi}\Big[\tfrac12 \big( \bpsi^\rh\Gamma^{ab}\gamma_\rh  -2{\bar\chi}\Gamma^{ab}\big) \psi_{\m\n} Q_{+}{}^{\m\n}{}_{ab}  + 4P_\n{}^{ab}\, \bpsi^{\m\n} \Gamma_a D_\m^{(sc)} \psi_b  
\nn\w2
& \qquad\quad   -Y^{\m\n} \bpsi^a \g_\m D_\n^{(sc)} \psi_a -3Z^{\m\n (ab)}\bpsi_a \gamma_\m D_\n^{(sc)} \psi_b
\nn\w2
& \qquad\quad
 -P^{\m c}{}_{[a} P^{\n d}{}_{b]} \bpsi^a \g^\m \Gamma^{cd}  D_\n^{(sc)} \psi^b 
 +2 \bpsi^\rh\gamma_\m\gamma_\rh \Gamma^b \big(D_\n^{(sc)} \psi^a\big) D^\m P^\n{}_{ba} 
\Big]
\w4
\cL_5 &= \alpha' e e^{2\varphi}\Big[-2( D_m^{(sc)} \bpsi^a ) \slashed{D} (\omega,\Omega_-) D_m^{(sc)} \psi_a +4{\bar\chi}\gamma_\m \Gamma^a (D_\n^{(sc)} \psi^b ) D^\m P^\n{}_{ab}
\nn\w2
&\qquad\quad -\tfrac12 Y \bpsi^a \slashed{D}^{(sc)}\psi_a +\tfrac12 Z_{ab} \bpsi^a \slashed{D}^{(sc)} \psi^b 
+\tfrac14 Q_{+\m\n cd} \bpsi^a \g^{\m\n}\Gamma^{cd}\slashed{D}^{(sc)} \psi_a 
\nn\w2
&\qquad\quad -\tfrac12 P_{\m ca} P_{\n db} \bpsi^a \Gamma^d\Gamma^c \g^{\m\n} \slashed{D}^{(sc)} \psi^b -\tfrac12 P_\m{}^c{}_{[a} P^{\m d}{}_{b]} \bpsi^a \Gamma_{cd} \slashed{D}^{(sc)} \psi^b 
\nn\w2
& \qquad\quad 
+P_\m{}^c{}_{(a} P^{\n d}{}_{b)}  \bpsi^a  \Gamma_{cd}  \g^\m D_\n^{(sc)} \psi^b  +\tfrac12 Q_{+}{}^{\m\n}{}_{cd} \bpsi^a \g_\m\Gamma^{cd}D_\n^{(sc)}\psi_a
 \nn\w2
& \qquad\quad 
+ \tfrac92 Q_-{}^{\m\n}{}_{ab}\bpsi^a \gamma_\m D_\n^{(sc)} \psi^b
+ \tfrac12 \bpsi^d \g^\m  \Gamma^{ab} \psi_d  P_{\m a}{}^c D^\n P_{\n bc}
\nn\w2
& \qquad\quad  -\tfrac12 \bpsi^a \g^\m  \Gamma^c \Gamma^d \psi^b P_{\m c b} D^\n P_{\n d a}  
\Big]
\w4
\cL_6 =& \alpha' e e^{2\varphi} H_{\m\n\rh}\Big[\tfrac{1}{12} Y_{\lambda\tau cd}\,\bpsi^b \Gamma^c\Gamma^d \gamma^\lambda \gamma^\tau \gamma^{\m\n\rh} \psi_b
 -\tfrac{1}{12} P_{\lambda cb} P_{\tau da}\,\bpsi^a \Gamma^c\Gamma^d \gamma^\lambda \gamma^\tau \gamma^{\m\n\rh} \psi^b
\nn\w2
& \qquad\quad + Q_{-}{}^{\m\n}{}_{ab}\,\bpsi^a \gamma^\rh \psi^b  -\tfrac16  \big(D^\s_{(sc)} \bpsi^a\big) \gamma^{\m\n\rh} D_\s^{(sc)} \psi_a 
\Big]\ ,
\end{align}}
where we have used the relation 
$ D_\m( \om, \Gamma_+ ) \g^\n = - H_\m{}^\n{}_\rh \g^\rh$, 
and the non-standard covariant derivatives are as defined in \eq{CDs} and %
\begin{align}
D_\m(\omega,\Omega_-) (D_m \psi_a) =&\, ( \pd_\m + \tfrac14 \omega_{\m pq} \gamma^{pq} + \tfrac14 Q_{+ \m b c} \Gamma^{bc} ) (D_m \psi_a) \ ,
\nn\w2 
& + \Omega_{- \m m}{}^n (D_n \psi_a) + Q_{-\m a}{}^b (D_m \psi_b) \ .
\end{align}
The structures that arise in the result for Lagrangian are grouped as follows. In the first term in \eq{OrgL}, with $\cL_0$ from \eq{6D}, only the zeroth and first order in $\alpha'$ that are to be kept. In the Lagrangian $\cL(R^2)$, the dependence on the hyperscalars enters only through the composite connections in the covariant derivatives. The Lagrangian $\cL_1$ contains the bosonic four derivative terms built out of hyperscalars. Denoting a generic fermion by $\psi$, the terms in $\cL_2$ schematically are of the form $\bpsi \psi\,(P DP)$, where the $PDP$ factor cannot be written as $D(PP)$. Thus there is no room for use of equations of motion here. Similarly, $\cL_3$ contains terms of the form $\bpsi \psi P^3$ with no room for equations of motion. The Lagrangian $\cL_4$ contains terms of the form $ P^2\, \bpsi D\psi $ or $(DP) \bpsi D\psi$. Terms in which the lowest order in $\alpha'$ field equations can arise directly or upon partial integration are collected in $\cL_5$, and the Lagrangian $\cL_6$ has the terms in which $H$ appears explicitly, as opposed to entering through covariant derivative as torsion. It is worth noting that many simplifications have occurred by working with the supercovariant derivative of the hyperino $D_\m^{(sc)}\psi_a$ defined in \eq{mdefs}. 

The supertransformations in terms of the redefined fields, and with prime notation dropped, are given by
\begin{align}
\delta e_\mu{}^r &= \bar{\epsilon} \gamma^r \psi_\mu\ , 
\nn\w2
\delta \psi_\mu &= D_\mu(\Omega_+) \epsilon  -\tfrac32 \alpha' \big[ \omega^L_{\m\n\rh}(\Omega_-) + \omega^Q_{\m\n\rh}(Q_-) \big] \gamma^{\n\rh} \e   - \alpha' P_{\n a}{}^c D_\m P^\n{}_{bc} \Gamma^{ab} \e\  , 
\nn\w2
\delta B_{\mu\nu} 
&= - \bar{\epsilon} \gamma_{[\mu} \psi_{\nu]} +2\alpha' \Omega_{-[\m}{}^{rs} \delta_0 \Omega^{(sc)}_{-\n] rs} +2\alpha' Q_{-[\m}{}^{ab} \delta_0 Q^{(sc)}_{-\n]ab}  \ , 
\nn\w2
\delta\chi 
&= \tfrac{1}{2} \gamma^\mu \epsilon \partial_\mu  \varphi 
- \tfrac{1}{12} H_{\mu\nu\rho} \gamma^{\mu\nu\rho} \epsilon  +\tfrac12 \alpha' \big[ \omega^L_{\m\n\rh}(\Omega_-) + \omega^Q_{\m\n\rh}(Q_-) \big] \gamma^{\m\n\rh} \e \ , 
\nn\w2
\delta\varphi &= \bar{\epsilon} \chi\ , 
\nn\w2
W \delta W^{-1} &
= \left(
\begin{array}{c|c}
 0 \  & 
 \ - \bar{\e} \Gamma_a \psi_b + 2 \al' \bar{\e} \Gamma_c \psi_b Y_a{}^c 
\\
\hline
\ - \bar{\e} \Gamma_b \psi_a + 2 \al' \bar{\e} \Gamma_c \psi_a Y_b{}^c   & 
0 \ 
\end{array}
\right)\ .
\nn\w2
\delta \psi_a 
&=   - \tfrac12 \g^\m \Gamma^b \e  P_{\m ba} - \al'\g^\m \Gamma^b \e  P_\m{}^c{}_a  Y_{bc}\ . 
\label{susy3}
\end{align}
It is understood that the quartic fermion terms in the action and the cubic fermion terms in the supertransformations are to be dropped. 

We find the commutation relation of these supertransformations as 
\begin{equation}
[ \delta_1, \delta_2 ] 
= \delta_{\rm g.c.}(\xi) + \delta_L(\lambda) 
+ \delta_{\rm tensor}(\Lambda) + \delta_{SO(4)_+}(\Lambda_+) 
+ \delta_{SO(4)_-}(\Lambda_-)\ , 
\label{commutator}
\end{equation}
where 
\begin{align}
\xi^\mu &= \bar{\epsilon}_2 \gamma^\mu \epsilon_1, 
\nn\w2
\lambda_{rs} &= - \xi^\mu \Omega_{-\mu rs} 
+ 6\alpha' \xi^\mu \left[ \omega^L_{\mu rs}(\Omega_-) 
+ \omega^Q_{\mu rs}(Q_-) \right]\ ,
\nn\w2
\Lambda_\mu &= \tfrac12 \xi_\m - \xi^\nu B_{\nu\mu}\ ,
\nn\w2
\Lambda_{\pm ab} &= \xi^\mu Q_{\pm \mu ab}\ .
\end{align}
To find supertransformation which may appear on the right-hand side 
of (\ref{commutator}) we need cubic fermi terms 
in the supertransformations.  In our conventions $\delta_L(\lambda) e_\mu{}^r= - \lambda^r{}_s e_\mu{}^s$. Note also the transformation rules $\delta B_{\m\n} = 2 \partial_{[\mu} \Lambda_{\nu]}$, , and those given in \eq{PT} and \eq{BT}. It is easy to check (\ref{commutator}) for $e_\mu{}^r$ and $\varphi$.  To check (\ref{commutator}) for $B_{\mu\nu}$ we used 
\begin{equation}
\delta_0 \Omega^{(sc)}_{-\mu rs} 
= -\bar{\epsilon} \gamma_\mu \psi_{rs}\ , \qquad
\delta_0 Q^{(sc)}_{-\mu ab}
= 2 P_{\mu c[a} \bar{\epsilon} \Gamma^c \psi_{b]} 
\end{equation}
and 
\begin{align}
[ \delta_{01}, \delta_{02}] \Omega^{(sc)}_{-\mu rs} 
&= - \xi^\nu R_{\mu\nu rs}(\Omega_-)\ ,
\nn\w2
[ \delta_{01}, \delta_{02}] Q^{(sc)}_{-\mu ab} 
&= - \xi^\nu Q_{-\mu\nu ab}\ .
\end{align}
To check (\ref{commutator}) for $W$ we used 
\begin{equation}
\delta_1 \left( - \bar{\e}_2 \Gamma_a \psi_b 
+ 2 \al' \bar{\e}_2 \Gamma_c \psi_b Y_a{}^c \right) 
- (1 \leftrightarrow 2)
= - \xi^\mu P_{\mu ab}
\end{equation}
and found 
\begin{align}
[ \delta_1, \delta_2] W^{-1}
&= W^{-1} \left( 
\begin{array}{cc}
0 & - \xi \cdot P \\
-\xi \cdot P^T & 0
\end{array}
\right) 
\nn\w2
&= \xi^\mu \partial_\mu W^{-1} 
- W^{-1} \left( 
\begin{array}{cc}
\xi \cdot Q_{+} & 0 \\
0 & \xi \cdot Q_{-}
\end{array}
\right)\ , 
\end{align}
where we have used (3.12) in the second line. 
We have not checked (\ref{commutator}) for fermi fields, 
which needs cubic fermi terms in the supertransformations. 

\end{appendix}

\end{document}